\documentclass[onecolumn,noshowpacs,nofootinbib,11pt]{revtex4}
\usepackage{float,xcolor,upgreek,yfonts,tikz}
\usepackage{color} 
\usepackage{textcomp,bm}
\usepackage{graphicx}
\usepackage{subfigure}
\usepackage{braket}
\usepackage{physics}
\usepackage[normalem]{ulem}

\def\b0{{\bf 0}}

\def\Re{\mathrm {Re}\;}
\def\bdm{\begin{displaymath}}
\def\edm{\end{displaymath}}

\def\Tr{{\mathrm{Tr}}}

\newcommand{\be}{\begin{eqnarray}}
\newcommand{\ee}{\end{eqnarray}}
\renewcommand{\b}{\mathring b}

\newcommand{\lp}{\ell_{\rm p}}
\newcommand{\mpl}{m_{\rm p}}
\newcommand{\gn}{G_{\rm N}}

\newcommand{\Det}{\text{Det}}
\graphicspath{{FinalVersionV1/FinalVersionV1}}
\usepackage{graphicx,float}\usepackage[all]{xy}
\usepackage{amsmath,upgreek,accents}
\definecolor{applegreen}{rgb}{0.55, 0.71, 0.0}

\definecolor{amber}{rgb}{1.0, 0.55, 0.1}

\usepackage{amssymb}

\usepackage{textgreek}
\newcommand{\beq}{\begin{eqnarray}}
\newcommand{\eeq}{\end{eqnarray}}

\newcommand{\bea}{\begin{eqnarray}}
\newcommand{\eea}{\end{eqnarray}}

\usepackage[T3,T1]{fontenc}
\usepackage[colorlinks,hyperindex,unicode]{hyperref}
\definecolor{green1}{RGB}{0,128,0} 
\hypersetup{hidelinks,backref=true,pagebackref=true,hyperindex=true,colorlinks=true,breaklinks=true,urlcolor= blue}
\hypersetup{%
  colorlinks = true,
  linkcolor  = blue,
  citecolor = cyan,
}
\usepackage{bookmark,textgreek}
\usepackage{hyperref,color,xcolor}
\newcommand\orcidroldao{{\href{https://orcid.org/0000-0003-3978-532X}{\orcidicon}}}
\newcommand\orcidkuntz{{\href{https://orcid.org/0000-0003-2621-5715}{\orcidicon}}}
\newcommand\orcidcasadio{{\href{https://orcid.org/0000-0002-1330-7787}{\orcidicon}}}
\newcommand{\orcidicon}{
	\begin{tikzpicture}
	\draw[lime, fill=lime] (0,0)
		circle [radius=0.16]
		node[white] {{\fontfamily{qag}\selectfont \tiny ID}};
	\draw[white, fill=white] (-0.0625,0.095)
		circle [radius=0.007];
	\end{tikzpicture}	\hspace{-2mm}
}
\begin{document}
\title{When gravitational decoupling and quantum gravity (re)unite}
\author{R.~Casadio
\orcidcasadio}
\affiliation{Dipartimento di Fisica e Astronomia, Universit\`a di Bologna, via Irnerio~46, 40126 Bologna, Italy}
\affiliation{I.N.F.N., Sezione di Bologna, I.S.~FLAG, viale B.~Pichat~6/2, 40127 Bologna, Italy}
\email{casadio@bo.infn.it}
\author{I.~Kuntz
\orcidkuntz}
\affiliation{Departamento de F\'isica, Universidade Federal do Paran\'a, PO Box 19044, Curitiba -- PR, 81531-980, Brazil.}
\email{kuntz@fisica.ufpr.br}
\author{R.~da~Rocha
\orcidroldao}
\affiliation{Federal University of ABC, Center of Mathematics,  Santo Andr\'e, 09210-580, Brazil.}
\email{roldao.rocha@ufabc.edu.br}
\begin{abstract}
The effective action for quantum gravity coupled to matter contains corrections arising from the
functional measure.
We analyse the effect of such corrections for anisotropic self-gravitating compact objects described
by means of the gravitational decoupling method applied to isotropic solutions of the Einstein field equations.
In particular, we consider the Tolman~IV solution of general relativity and show that quantum gravity effects
can modify the effective energy density as well as the effective tangential and radial pressures. For a suitable choice of the mimicking constant, 
 upper bounds on the quantum corrections can be driven by the surface redshift of the anisotropic compact stellar system obtained with the gravitational decoupling. 
\end{abstract}
\maketitle
\section{Introduction}
\label{s1}
The experimental detection of gravitational waves radiated from the final stages
of binary mergers~\cite{LIGOScientific:2016lio} has opened a window into the
strong gravity regime.
Solutions of the Einstein field equations and their generalisations can
therefore be experimentally tested.
The gravitational decoupling (GD) of the Einstein equations is a method for 
obtaining self-gravitating compact stellar configurations starting from 
known solutions of general relativity (GR).
Anisotropic stars arise in a very natural way in the GD approach, yielding
the possibility of obtaining analytical solutions to the Einstein field equations
supplemented by general forms of the energy-momentum
tensor~\cite{Ovalle:2017fgl,Ovalle:2019qyi,Ovalle:2017wqi}. 
The GD includes, as a particular case, the minimal geometrical deformation
(MGD)~\cite{Ovalle:2007bn,Ovalle:2018ans}, which was originally formulated
in order to describe compact stars and black holes in the
brane-world~\cite{Antoniadis:1998ig,daRocha:2012pt,Abdalla:2009pg}, including
soft hair~\cite{Ferreira-Martins:2021cga}. 
\par
In the GD method, sources of the GR gravitational field and the corresponding 
field equations are split into two parts.
The first one describes a GR solution, whereas the second part contains 
additional sources, which can carry any type of charge, including tidal and
gauge ones, hairy fields of any physical origin, as well as any other contributions
from extended models of gravity.
Examples of configurations so obtained can be found in Refs.~\cite{Sharif:2023upz,
Estrada:2019aeh,Gabbanelli:2019txr,Leon:2023nbj,Avalos:2023ywb,
Contreras:2022vec,Maurya:2021mqx,Singh:2021iwv,
Maurya:2021tca,Maurya:2023szc,Singh:2020bdv,Ramos:2021drk,
Casadio:2019usg,Sharif:2023ecm,Rincon:2019jal,Tello-Ortiz:2019gcl,Morales:2018urp,
Panotopoulos:2018law,Singh:2019ktp,Jasim:2023ehu}.
Realistic models based on relativistic description of nuclear interactions
suggest that star interiors are anisotropic at extremely high densities.
The GD method easily allows for including pressure
anisotropies~\cite{Gabbanelli:2018bhs,PerezGraterol:2018eut,
Heras:2018cpz,Torres:2019mee,Hensh:2019rtb,Contreras:2019iwm,
Tello-Ortiz:2021kxg,Sharif:2020vvk,Contreras:2019mhf,
Andrade:2023wux,Zubair:2023cvu,Bamba:2023wok,Maurya:2023uiy,
Tello-Ortiz:2023poi,Contreras:2021xkf,Sharif:2020lbt,Maurya:2020ebd,
Ovalle:2020kpd,Ovalle:2021jzf,Contreras:2021yxe}.
The MGD applies naturally in the AdS/CFT scenario~\cite{daRocha:2020jdj,
DaRocha:2019fjr,daRocha:2020gee} for studying black holes
with physically viable low-energy limits~\cite{ssm1,Shiromizu:2001jm}.
The trace and Weyl anomalies were also calculated for hairy GD solutions,
establishing new possibilities of employing the AdS/CFT to the membrane
paradigm~\cite{Meert:2020sqv,Meert:2021khi}.
Quasinormal modes radiated from hairy GD solutions were also addressed
recently in Ref.~~\cite{Cavalcanti:2022cga,Yang:2022ifo,Li:2022hkq,Avalos:2023jeh,Al-Badawi:2024iax}. 
\par
In the functional approach to quantum gravity, a functional measure
is required for obtaining an effective action which remains invariant under field
redefinitions (hence gauge transformations, see ~\cite{Parker:2009uva} and
references therein).
This approach was employed in previous works
to study the spacetime stability~\cite{Kuntz:2019gup} as well as 
the weakly-coupled gravity and the strongly-coupled conformal field theory
sides of the gauge/gravity correspondence~\cite{Kuntz:2019omq}.
Corrections to transport and response coefficients in relativistic second-order
hydrodynamics were obtained using the linear response procedure.
The shear viscosity, entropy density, diffusion constant, and speed of sound, 
were shown not to achieve any corrections from the functional measure of gravity. 
On the other hand, the energy density, pressure, relaxation time, shear relaxation,
bulk viscosity, decay rate of sound waves, and coefficients of conformal traceless
tensor fields, were shown to carry significant quantum corrections due to the
functional measure, also reflecting the instability of the strongly-coupled fluids
on the boundary CFT.
This opens up the possibility of testing quantum gravity with the quark-gluon plasma,
as neutron stars can contain hadronic and quark phases.
Analogously to asymptotic freedom, which permits matter deconfinement when
the density increases at low temperatures, this kind of phenomenon could
naturally occur in the inner core of neutron stars and quark
stars~\cite{Sharif:2022vvi,Tello-Ortiz:2020euy,Maurya:2022hmw}. 
\par
The main goal of this paper is to study the effects of the contribution from the
functional measure in the effective action for anisotropic stellar configurations
obtained with the GD method in general.
In particular, we will then consider the MGD of the Tolman~IV solution found in
Ref.~\cite{Ovalle:2017wqi} to include quantum gravity effects analysed in
Ref.~\cite{deFreitas:2023ujo}. 

We will deal with complex metric functions by simply considering their modulus
and the Kontsevich--Segal (KS) condition that complex metrics should satisfy
in order to represent acceptable backgrounds for quantum field theory (QFT)~\cite{Kontsevich:2021dmb}.  
In particular, the KS theorem states that the sum of the modulus of the arguments
of the eigenvalues of a complex metric must be less than $\pi$, which we will
take as a criterion hereby. 
We will show that instabilities generated by the functional measure can be
cancelled and the metric components remain real in the interior of the stellar
distribution for some choice of the mimic constraint. \textcolor{black}{Our results shed new light on complex spacetime geometries compatible with a path integral formulation of quantum gravity, as recently addressed by Ref. \cite{Liu:2024zti}, also in the context of quantum aspects of black holes.}
However, the effective energy density still carries instabilities of the degrees
of freedom in the fluid, and the effective tangential and radial pressures are
affected by quantum gravity effects. For another choice of the mimic constraint, the ADM mass, the effective radial and tangential pressures as well as the energy densities carry effects of both the GD hairy charge and the parameter regulating quantum gravity effects. 
\par
The paper is organized as follows:
in Section~\ref{measure1} the functional measure is introduced in quantum field theory
(QFT), playing an important role in the construction of the configuration-space metric,
contributing to an additional expression in the effective action of quantum gravity
corresponding to 1-loop corrections.
Section~\ref{s2} is devoted to briefly reviewing the GD setup and analyzing
modifications due to the functional measure.
Compact self-gravitating stellar systems are scrutinized in this context.
Section~\ref{mms5} addresses the GD of Tolman~IV solutions with quantum gravity
corrections. The surface redshift bounds are then used to constrain the maximum magnitude of the parameter governing quantum gravity effects. 
Section~\ref{secc} presents some concluding remarks and future perspectives. Appendixes (\ref{app1}, \ref{app2}) present other choices of the mimicking constraints and subsequent analyses. 
\section{Functional measure in quantum field theory}
\label{measure1}
Despite the wide applicability of path integrals in physics, a well-defined mathematical
construction, and whether it can be indeed interpreted as an integral, remains unknown.
To a great extent, the problem boils down to the definition of the functional measure.
The issue has been studied since long and different definitions have been
suggested~\cite{Unz:1985wq,Toms:1986sh,Moretti:1997qn,Hatsuda:1989qy,
vanNieuwenhuizen:1989dx,Armendariz-Picon:2014xda,Becker:2020mjl,
Buchbinder:1987vp,Hamamoto:2000ab}.
Although still quite formal, such manipulations could have phenomenological
consequences that have largely been dismissed by the use of dimensional regularization,
where the measure is regularized to unity.
\par
Dimensional regularization is indeed the most prominent form of regularization
used nowadays when dealing with field theories in the continuum.
It has several advantages, including the ease-of-use and preservation of symmetries
at every step of the calculation~\cite{tHooft:1972tcz}.
However, applying dimensional regularization to the functional measure only hides the issue.
Anomalies, for example, result from the non-invariance of the functional measure,
which could have never been observed if dimensional regularization were adopted
in the path integral.
\par
Other forms of regularization, such as cutoff and lattice, forces one to deal with
all aspects of the functional measure.
In particular, in the absence of a rigorous definition of the path integral,
its interpretation relies on the continuum limit of a lattice.
This procedure-based approach, albeit ambiguous, is the only known way
of giving meaning to the Feynman integrals and necessarily requires
understanding of the functional measure.
The existence of physical cutoffs, as implemented in Wilson's effective
theory~\cite{Wilson:1974mb} and envisaged in minimal length
scenarios~\cite{Hossenfelder:2012jw,Casadio:2020hzs,Casadio:2022opg},
is yet another reason that calls for a better understanding of the integration measure. 
\par
From a geometrical perspective, invariance under all of the underlying symmetries
is obtained from the generating functional
\begin{equation}
Z[J]
=
\int\dd\mu[\varphi]\,
e^{\frac{i}{\hbar}\left( S[\varphi] + J_i\, \varphi^i\right)}
\ ,
\label{Z}
\end{equation}
where $S[\varphi]$ is the bare action, $\varphi^i = (\phi(x), A_\mu(x), g_{\mu\nu}(x), \ldots)$
denotes a set of fields (formally) sourced by the external currents $J_i$, and
\begin{equation}
\dd\mu[\varphi]
=
\prod_i \dd\varphi^i\,\sqrt{\Det\, G_{jk}(\varphi)}
\ ,
\label{measure}
\end{equation}
is the functional measure.
$\Det \,G_{ij}(\varphi)$ denotes the functional determinant of the field configuration-space
metric $G_{ij}$.
The factor $\sqrt{\Det\, G_{ij}}$ must be included to cancel the Jacobian from
$\prod_i \mathrm{d}\varphi^i$, thus leaving the total measure $\mathrm{d}\mu[\varphi]$ invariant
under field redefinitions.
One should note that such a factor is required even for a flat configuration space,
because the Jacobian of general field transformations is not one. 
Using the relation $\Det\log = \log\Tr$, we can write
\begin{align}
\Det \,G_{ij}
=
e^{\int\dd^4x \,\sqrt{-g} \, \tr\log G_{IJ}}
\ ,
\label{detG}
\end{align}
where $g$ is the determinant of the spacetime metric $g_{\mu\nu}$ and we
used the functional trace
\begin{equation}
\Tr\, A_{ij}
=
\int\dd^4x \,\sqrt{-g} \, \tr A_{IJ}(x,x)
\ ,
\label{trace}
\end{equation}
including summation over discrete indices ($I$, $J$), via the ordinary trace $\tr$,
and integration over spacetime.
From Eqs.~\eqref{Z} and \eqref{detG}, one then finds 
\begin{equation}
Z[J]
=
\int \prod_i \dd\varphi^i
\,e^{\frac{i}{\hbar}\left( S_\text{eff}[\varphi] + J_j\, \varphi^j\right)}
\ ,
\end{equation}
where we defined the Wilsonian effective action~\footnote{It is clear that the 
contribution from the functional measure is of the same order (in $\hbar$)
as 1-loop corrections which we will discuss later.}
\begin{equation}
S_G
=
\int\dd^4x\, \sqrt{-g}
\left(
\mathcal L
-
\frac{i}{2}\,\hbar\,\tr\log G_{IJ}
\right)
\ ,
\label{genact}
\end{equation}
for some bare Lagrangian $\mathcal L$ corresponding to the bare action $S$.
\par
The configuration-space metric must be seen as part of the definition of the theory,
hence physical systems are now fully described by the pair $(S[\varphi], G_{ij})$.
The sole specification of the classical action can no longer secure uniqueness,
with different $G_{ij}$ representing different quantization schemes.
The determination of $G_{ij}$ follows closely the procedure to obtain the action.
As often done, we shall assume ultralocality, namely
\begin{equation}
G_{ij}
=
G_{IJ}(\varphi)\, \delta^{(4)}(x,x')
\ ,
\label{ultralocal}
\end{equation}
and some symmetry principles to fix $G_{IJ}(\varphi)$.
Note that $G_{IJ}(\varphi)$ is a function (not functional) of the fields
which describes the metric of the finite-dimensional space obtained
by fixing the spacetime point.
The form of the configuration-space metric in Eq.~\eqref{ultralocal}
guarantees the same $G_{IJ}$ across all spacetime points and prevents
interactions from distant events. 
\par
From a physical viewpoint, there is nothing fundamental about ultralocality.
One could as well have assumed locality instead, in which case there
would appear terms with derivatives of the Dirac delta in Eq.~\eqref{ultralocal}, {\color{black}or even non-locality, where there would be contributions from different spacetime points.
However, when the configuration-space metric is identified from the kinetic term \cite{Vilkovisky:1984st,Alonso:2015fsp,Alonso:2016oah,Finn:2019aip}:
    \begin{align*}
    S
    &=
    \int\mathrm{d}^4x \, G_{IJ}(\varphi) \, \partial_\mu \varphi^I(x) \partial^\mu \varphi^J(x)
    \\
    &=
    \int\int\mathrm{d}^4x \, \mathrm{d}^4x' \, G_{IJ}(\varphi) \, \delta^{(4)}(x,x') \, \partial_\mu \varphi^I(x) \partial^\mu \varphi^J(x')
    \ ,
    \end{align*}
    thus $G_{ij} = G_{IJ}(\varphi) \, \delta^{(4)}(x,x')$.
	Ultralocality then only reflects the locality of the Lagrangian. Even when the configuration-space metric is not defined via the action as above \cite{Kuntz:2024opj,Berganholi:2025hmi,deFreitas:2023ujo,Kuntz:2022kcw}, one adopts ultralocality as a simplifying assumption as it is a lot easier to deal with logarithms of ultralocal quantities than non-local ones.
}

In fact, the ultralocality assumption allows one to make sense of the
functional logarithm as the continuum extension of the logarithm of a tensor product as we explain in the following.
{
\color{black}
First we recall that $i = (I, x)$ contains discrete indices $I$ and continuum ones $x$. Hence, the term $G_{IJ}(\varphi) \, \delta^{(4)}(x,x')$ is the component of a tensor product. The Dirac delta plays the role of coordinates of the identity $\mathbb{I}$ in the position basis. Hence one could write the configuration-space metric as $G \otimes \mathbb{I}$, where $G = G_{IJ} \, \mathrm{d}\varphi^I \otimes \mathrm{d}\varphi^J$. For linear operators $A$ and $B$, the well-known formula
    \begin{equation}
    \log(A \otimes B)
    =
    \log(A) \otimes \mathbb{I}_B + \mathbb{I}_A \otimes \log(B) 
    \end{equation}
    can be applied to wit:
    \begin{equation}
    \log(G \otimes \mathbb{I})
    =
    \log(G) \otimes \mathbb{I}
    \ ,
    \end{equation}
    or in coordinates:
    \begin{equation}
		\log\left[G_{IJ}\, \delta^{(4)}(x,x')\right]
		=
		\log(G_{IJ}) \,\delta^{(4)}(x,x')
		\ .
		\label{dsum}
	\end{equation}
    This justifies the well-spread pull-out of the Dirac delta from the logarithm in quantum field theory.
}

However, ultralocality causes the appearance of $\delta^{(4)}(0)$ in spacetime
integrations, as can be seen from Eqs.~\eqref{trace} and \eqref{dsum}.
Replacing dimensional regularization in favour of other regulators is a way to explore
functional measure effects without facing the difficulties introduced by trading ultralocality
for the locality.
\par
We shall here adopt a Gaussian regularization
\begin{equation}
\delta^{(4)}(x)
\to
\frac{e^{-\frac{x^2}{2\,L_{\rm UV}^2}}}
{(2\,\pi)^{2}\,L_{\rm UV}^4}\,
\ ,
\label{Greg}
\end{equation}
where $K_{\rm UV}=\hbar/L_{\rm UV}$ is a Wilsonian UV cutoff. $K_{\rm UV}$ is assumed to be much larger than the scales of interest, so that it only plays a role in the energy expansion, but momenta integrals are not hard cutoff.
Note that while the Gaussian representation of $\delta^{(4)}(x)$ depends on the sign of $x^2$ (hence on the chosen frame), the regularization of the $\delta^{(4)}(0)$ divergence does not. One finds $\delta^{(4)}(0) = L_{\rm UV}^{-4} / (2\pi)^2$ for any frame because $\delta^{(4)}(x)$ is evaluated at $x^2 = 0$.

{\color{black}
One could worry that the regularization \eqref{Greg} could introduce non-local couplings, seemingly contradicting the ultralocal assumption. However, this is typical in effective field theory: the effective action can become non-local even for an initial local classical action. There is no conceptual tension because the non-locality is not fundamental. It arises from the way the effective action represents integrated-out modes with new effective interactions, but the theory remains local at the regime where the effective theory is applicable. Indeed, for energies $E \ll K_{\rm UV}$, the apparent non-locality is reorganized into an infinite tower of local interactions, capturing UV effects via local operators. The same rationale applies in the presence of an ultralocal measure. We stress that the resulting effective action (see Eq.~\eqref{eq:newac}) is local. This is expected since the functional measure is a UV object tied to the $\delta^{(4)}(0)$-divergence.
}

In studying compact objects of mass $M$ and size $R_{\rm s}$, we shall be typically interested
in configurations with energy densities $\rho\sim M/R_{\rm s}^3\ll \mpl/\lp^3$, where $\mpl$
is the Planck mass and $\lp$ the Planck length.~\footnote{In units with the speed of 
light $c=1$, the (reduced) Planck mass and length are given by $\gn=\lp/\mpl$
and $\hbar=\mpl\,\lp$.
In the following, we shall also use $\kappa=8\,\pi\,\gn$ and the spacetime signature
$(-+++)$.}
We can therefore assume $\rho\ll \hbar/L_{\rm UV}^4 \ll \mpl/\lp^3$
(or, typically, $\lp\ll L_{\rm UV}\ll R$). 
This way, we obtain
\begin{equation}
S_\mu
=
\int\dd^4x\, \sqrt{-g}
\left(
\mathcal L
-
i\, \zeta \, \tr\log G_{IJ}
\right)
\ ,
\label{genact}
\end{equation}
where $\zeta=\zeta(K_{\rm UV})\propto \hbar/L_{\rm UV}^4$ is a Wilsonian coefficient
whose running is such that
\begin{equation}
K_{\rm UV}\,
\frac{\dd Z[J]}{\dd K_{\rm UV}}
=
0
\ .
\label{RGE}
\end{equation}
Note that the correction due to the measure scales quartically with the cutoff $K_{\rm UV}$,
thus $\tr\log G_{IJ}$ is a relevant operator and $\zeta$ is UV sensitive.

Note that the functional measure contributes with an imaginary term in Eq.~\eqref{genact}, which ultimately {\color{black} would} lead to complex quantities.  
There is nothing worrisome about complex quantities, as long as one imposes some reality condition to be satisfied by physical observables.

Indeed, auxiliar quantities that show up at intermediate steps of the calculation, such as the effective action or n-point functions, need not be real as they are not observables. They are, nonetheless, related to physical observables by some realization procedure, such as taking the modulus squared in the case of probabilities.

We must stress, however, that there is no clear-cut definition of observables anywhere in physics. It is up to us to define the procedure to connect theory with experiment and hence make predictions. Naturally, choices to implement the reality condition other than the absolute value might exist and lead to different predictions. {\color{black} However, the absolute value follows from very minimal and reasonable assumptions (see Appendix \ref{app:abs}).} At the end of the day, only observations will tell what prescription shall reign.

We shall implement the reality condition by taking the modulus of the metric. {\color{black} We shall therefore define ``physical quantities'' by the absolute value of the complex ones.} Although this procedure might be struck as ad-hoc, it is no different than
{\color{black}defining observables via $|\psi(x)|^2$ of the complex-valued wavefunction $\psi(x)$.
As mentioned above, taking the absolute value is an additional and non-unique prescription. In order to link complex quantities with experiments, a choice of some reality condition must be made. Taking either the real or imaginary part would dismiss part of the contribution, in addition to  being a coordinate-dependent definition since they are Cartesian coordinates of the complex plane. Under the assumptions of coordinate-independence and scaling homogeneity, the absolute value is the unique reality condition (see Appendix \ref{app:abs}).
Therefore, the most obvious choice would be to take the modulus, which
}
is a legitimate procedure (and the simplest one) to connect theory with experiment. Furthermore, taking the modulus of a tensor component in one frame guarantees that such a quantity remains real in any frame because diffeomorphisms are real.

At the level of the effective action, imaginary terms correspond to vacuum instabilities.
These instabilities are easily interpreted in light of Schwinger's pair production, where the vacuum (or, more precisely, the background field) decays into pairs of particle and antiparticle \cite{Schwinger:1951nm} (see also Refs.~\cite{Dunne:2004nc,Dunne:2012vv,Gelis:2015kya} for reviews and \cite{Wondrak:2023zdi} for a recent example in the gravitational case). Since by definition, the effective action gives:
\begin{equation}
\langle out|in \rangle = e^{i \Gamma[g_\text{sol}] },
\end{equation}
where $g_\text{sol}$ is some solution to the equations of motion, a straightforward calculation shows that the vacuum persistance probabiliy reads:
\begin{equation}
|\langle out|in \rangle|^2 = e^{-2 \text{Im}(\Gamma[g_\text{sol}])}
\ .
\end{equation}

This instability corresponds to the vacuum decay into particles present in the fundamental theory. In our case, calculating precisely the decay products in general are clearly very difficult as this requires the knowledge of all degrees of freedom present in the bare Lagrangian. This is impossible to achieve in an bottom-up effective approach, such as the one adopted in this paper. In the simplest scenario, where matter fields are absent, the vacuum (background metric) decays into a pair of gravitons (the graviton is its own antiparticle).

From the aforementioned effective viewpoint, the vacuum decay is perceived as an exponential decay for the fields appearing in the effective action. Indeed, the imaginary term modifies the pole structure of the dressed propagators, yielding complex poles. The corresponding modes thus evolve with $\sim e^{ikx}$ with $k$ complex with positive imaginary part \cite{deFreitas:2023ujo}.

Also, it should be no surprise that the (vacuum expectation value~\footnote{We recall that the effective action is a functional of the mean fields, so in truth $\Gamma[\langle g \rangle]$. The corresponding equations of motion then describe $\langle g \rangle$, which can very well be complex.} of the) metric acquires an imaginary part. The same generally happens for any other field after quantization. This clearly requires a reinterpretation of classical objects, as has happened in, say, QED,  where the real part of the effective action accounts for the vacuum polarization and the imaginary one for the pair production. In quantum gravity, the spacetime is only instrumental for the definition of the theory, but not a physical observable itself.  Indeed, except for the case of real scalar fields, field operators are not Hermitian in general. Only observables should be real.
\par
Beside the functional measure discussed above, after performing the path integral in Eq.~\eqref{Z}
at the 1-loop level, the effective action also gains contribution from the Hessian
\begin{equation}
\Gamma[\varphi]
=
S_G[\varphi]
+
\frac{i}{2}\,\hbar\,\Tr\log \mathcal{H}_{ij}
\ .
\label{1loop}
\end{equation}
Note that, individually, the corrections in Eqs.~\eqref{genact} and \eqref{1loop} do not transform
covariantly, because the determinant of a 2-rank tensor is basis-dependent.
Their combination however results in 
\be
\Tr\log (G^{ik}\,\mathcal{H}_{kj}) = \Tr\log \mathcal{H}^i_{\ j}
\  ,
\label{TLGij}
\ee
which is invariant and shows the important role played by the functional measure.
\subsection{Effective action for gravity and matter}
For pure gravity, one has $\varphi^i = g^{\mu\nu}(x)$ and $i=(\mu\nu, x)$.
To lowest order, one then finds the DeWitt metric $G_{IJ} = G_{\mu\nu\rho\sigma}$,
where~\footnote{In principle, Eq.~\eqref{DWmetric} could be multiplied by a global factor
$g^\epsilon = (\det g_{\mu\nu})^\epsilon$.
We take $\epsilon=0$ for simplicity.}
\begin{equation}
G_{\mu\nu\rho\sigma}
=
\frac12
\left(
g_{\mu\rho}\, g_{\nu\sigma}
+
g_{\mu\sigma}\, g_{\nu\rho}
-
a \, g_{\mu\nu}\, g_{\rho\sigma}
\right)
\label{DWmetric}
\end{equation}
and $a$ is a dimensionless parameter.
\par
When matter is present, there could be other contributions to $G_{IJ}$.
For example, for a scalar field coupled to gravity $\varphi^i = (g_{\mu\nu}(x), \phi(x))$,
 the simplest non-trivial choice for the field-space metric would
read~\cite{Casadio:2021rwj,Kuntz:2022kcw}
\begin{equation}
G_{IJ}
=
\begin{pmatrix}
G_{\phi\phi} & 0 \\
0 & G_{\mu\nu\rho\sigma}
\end{pmatrix}
\ ,
\end{equation}
where
\begin{equation}
G_{\phi\phi}
=
c_1 + c_2\,L_{\rm UV}\,\frac{\phi}{\sqrt{\hbar}}
\ ,
\label{eq:metricscalar}
\end{equation}
and $c_i$ are free dimensionless parameters.
{\color{black}
We stress that there is no modelling around the configuration-space metric (and hence the functional measure). The form of the configuration-space metric is not cooked up, but it is rather determined using symmetry principles and written as an expansion in inverse powers of some physical cutoff, as it is typical in effective field theory. As a first approximation, we have kept only the leading order correction, which results in the DeWitt metric for gravity (Eq.~\eqref{DWmetric}) and Eq.~\eqref{eq:metricscalar} for the scalar sector.
}
{\color{black}
However, one could go to higher orders in $\phi$, or even write $G_{\phi\phi}$ as some non-polynomial function of $\phi$. Since the measure corresponding to $G_{\phi\phi}$ contributes as an effective potential to the effective action (see Eq.~\eqref{genact}), one could assume that $G_{\phi\phi}$ depends on the classical potential $V(\phi)$ via~\footnote{\color{black} We have included the constants $c_i$ in the definition of $\Upsilon$. Because of the logarithm in Eq.~\eqref{genact}, $c_1$ can be pulled out into a constant contribution to the effective action, which does not affect the equations of motion.}:
\begin{equation}
	G_{\phi\phi}
	=
	1
	+
	\frac{V(\phi)}{\Upsilon}
	\ ,
	\label{gpot}
\end{equation}
for some cutoff scale $\Upsilon \sim (\sqrt{\hbar}/L_{UV})^4$. The form of Eq.~\eqref{gpot} guarantees that $G_{\phi\phi}$ is a perturbative expansion around the trivial configuration-space metric.
For a stationary field, one could thus take:
\begin{equation}
	G_{\phi\phi}
	=
	1
	+
	\frac{\rho(\phi)}{\Upsilon}
	\ ,
	\label{grho}
\end{equation}
with $\rho(\phi)$ denoting the energy density. The same rationale could actually be applied to more general forms of matter fields, where in place of the scalar energy density one would find the energy density for generic fields $\rho=\rho(\varphi^I)$. The only caveat is that $\rho(\varphi^I)$ must be a scalar. This is the case for the electromagnetic field (though it is zero in the stationary regime), for complex scalar fields, for which $\rho(\varphi^I) = \rho(\Phi^\dagger \Phi)$, and for spinors, with $\rho(\varphi^I) = \rho(\Psi^\dagger \Psi)$. On the other hand, the energy density of massive vector fields, namely $\rho(\varphi^I) \sim m^2 \left(A_0^2	+\tfrac12 A_i^2 \right)$,  does not transform as a scalar, thus it cannot be accounted by our formalism. We shall thus consider the following general configuration-space metric for the matter sector:
\begin{equation}
	G_{\varphi\varphi} = 1 + \frac{\rho(\varphi)}{\Upsilon},
\end{equation}
for generic matter fields $\varphi^I(x)$.
}

Since
\begin{equation}
\det G_{IJ}
=
{\color{black}
\left(\det G_{\varphi\varphi}\right)
}
\left(\det G_{\mu\nu\rho\sigma}\right)
\ ,
\end{equation}
the measure term in Eq.~\eqref{1loop} splits into matter and gravity contributions
\begin{equation}
\Tr\log G_{ij}
=
\zeta
\int\dd^4x\, \sqrt{-g}
\left[
{\color{black}
\log(1 + \frac{\rho(\varphi)}{\Upsilon})
}
+ \tr\log G_{\mu\nu\rho\sigma}
\right]
\ .
\end{equation}
We have here determined $G_{\varphi\varphi}$ in the spirit of effective field theory by implementing
an expansion in the energy (density).
However, a fundamental measure could in principle be any function $f(\varphi)$ of the scalar field,
and we shall take advantage of this feature in the following.
\par
We shall take the bare Lagrangian for GR minimally coupled to matter, namely
\begin{equation}
\mathcal L
=
\frac{R}{2\,\kappa}
+
\mathcal L_{\rm m}(\varphi)
\ ,
\end{equation}
where $R$ is the Ricci scalar and $\mathcal{L}_{\rm m}$ is the Lagrangian for the
scalar field.
Then the Hessian reads
\begin{equation}
\mathcal{H}_{\mu\nu\rho\sigma}
=
K_{\mu\nu\rho\sigma}\,
\Box
+ U_{\mu\nu\rho\sigma}
\ ,
\label{hessianGR}
\end{equation}
where
\begin{equation}
K_{\mu\nu\rho\sigma} = \frac14
\left(
g_{\mu\rho}\, g_{\nu\sigma}
+
g_{\mu\sigma}\, g_{\nu\rho}
-
g_{\mu\nu}\, g_{\rho\sigma}
\right)
\end{equation}
and $U_{\mu\nu\rho\sigma}$ is a tensor that depends on the spacetime curvature,
whose form is not important.
From Eqs.~\eqref{DWmetric} and \eqref{hessianGR}, we can write~\eqref{TLGij}
as~\cite{deFreitas:2023ujo}
\begin{eqnarray}
\Tr\log \mathcal{H}^i_{\ j}
&\!=\!&
-\zeta
\int\dd^4x\,
\sqrt{-g}
\left\{
{\color{black}
\log(1 + \frac{\rho(\varphi)}{\Upsilon})
}
-
\log\det
[\frac12
\left(
\delta_{( \mu}^{\ \ \rho} \,\delta_{\nu )}^{\ \ \sigma}
+
(a-1)\,g_{\mu\nu}\,g^{\rho\sigma}
\right)]
\right.
\nonumber
\\
&&
\phantom{\zeta\int\dd^4x\,\sqrt{-g}\left\{\right.}
\left.
-
\log\det
[\delta_{( \alpha}^{\ \ \mu}\, \delta_{\beta )}^{\ \ \nu}\,
\Box
+
(K^{-1})^{\mu\nu\rho\sigma}\,U_{\rho\sigma\alpha\beta}]
\right\}
\ ,
\label{1piac}
\end{eqnarray}
where we pulled out $K_{\mu\nu\rho\sigma}$ from $H_{\mu\nu\rho\sigma}$
and put it along with $G_{\mu\nu\rho\sigma}$.
The last term in Eq.~\eqref{1piac} can be obtained as a power series in the curvature
or derivatives~\cite{DeWitt:2003pm,Barvinsky:1987uw,Barvinsky:1985an}.
Such a term represents next-to-leading order contributions when compared
to the second term in Eq.~\eqref{1piac}, which contains no derivatives,
thus we can safely drop it at low energies.
Finally, the matrix determinant lemma yields the effective action
\begin{equation}
\Gamma[g]
=
\int\dd^4x\, \sqrt{-g}
\left[
\frac{R}{2\,\kappa}
+
\mathcal{L}_m
+
i\, \frac{\zeta}{2}\,\log(\frac{4\,a-3}{256})
-
i\,\frac{\zeta}{2}\,\log(1 + \frac{\rho(\varphi)}{\Upsilon})
\right]
\ ,
\label{eq:newac}
\end{equation}
which depends on the parameters $\zeta\sim K_{\rm UV}^4$, $a$ and $\Upsilon$,
and on the (expectation value of the) scalar (matter) field in the configurations
of interest. The corresponding equations of motion read:
\begin{equation} 	
	G_{\mu\nu} 
	= 
	T_{\mu\nu}
	- \frac{i\zeta}{2}\log(\frac{4\,a-3}{256}) g_{\mu\nu}
	+ {\frac{i \zeta}{2} \log(1 + \frac{\rho(\varphi)}{\Upsilon}) g_{\mu\nu} }
	\ ,
	\label{eommm}
\end{equation} 
where $T_{\mu\nu}$ is the energy-momentum tensor for $\mathcal{L}_m$.

{\color{black}
One should note that no complexification has taken place in obtaining Eq.~\eqref{eommm}. No analytic continuation to the complex space was performed, hence there is no obvious way to return to the reals in the end. The imaginary term is present from the onset. This imaginary correction is forced upon us because of the functional measure in the path integral. Although the measure is real, the correction to the action becomes imaginary due to the $i = \sqrt{-1}$ in the argument of the exponent in the path integral. When one writes the measure as a correction to the action, the imaginary term shows up.
}
\subsection{Effective field equations for compact objects}
We observe that the term associated with the quantum measure for gravity
in Eq.~\eqref{eq:newac} is constant and would therefore be the same inside as
well as in the vacuum outside a compact matter source.
{\color{black}
On the other hand, the last term in Eq.~\eqref{eq:newac}, which corresponds to the matter contribution from the measure and depends on $\phi$, is not. Although both these terms could be about the same magnitude in the vacuum, the $\phi$-dependent term increases substantially inside a compact matter source while the $a$-dependent term remains constant, thus the former dominates. Since we are interested in the inside of a compact source,
it is safe to assume that it is negligibly small in comparison with the other terms.
}
The second correction depends on the {energy density},
and we may also assume that it is negligible in the vacuum where {\color{black} $\varphi^I\simeq 0$}.
\par
%Moreover, in order to consider more general forms of matter than just a scalar field,
%we notice that $\phi\sim \rho$, the (proper) energy density inside a compact object.
We can therefore write the effective field equations as
\begin{equation} 	
R_{\mu\nu}-\frac{1}{2}\,R\, g_{\mu\nu} 
= 
\kappa\,T_{\mu\nu}
-
i\,\kappa\,\frac{\zeta}{2}\,
\log(1 + \frac{\rho}{\Upsilon})\,
g_{\mu\nu}
\ ,
\label{eom}
\end{equation} 
where $T_{\mu\nu}$ is the energy-momentum tensor derived from
the matter Lagrangian $\mathcal{L}_m$
Clearly, GR is smoothly recovered in the limit $\zeta\to 0$, as
well as for $\Upsilon\to\infty$.
In fact, given that the effective action~\eqref{eq:newac} holds in the regime in which 
$\rho\ll\Upsilon$, we can further expand the quantum correction and
obtain
\begin{equation} 	
R_{\mu\nu}-\frac{1}{2}\,R\, g_{\mu\nu} 
\simeq 
\kappa\,T_{\mu\nu}
-
i\,\kappa\,\bar\zeta\,\rho\,
g_{\mu\nu}
\ ,
\label{eom1}
\end{equation}
where $\bar\zeta\simeq \zeta/2\,\Upsilon$ is a dimensionless parameter.
\par
The imaginary contribution in Eq.~\eqref{eom1} may yield complex solutions,
which requires some comments. The imaginary part of the effective energy-momentum in the effective Einstein's equations \eqref{eom1} is a consequence of using a quantum measure in the effective action, indicating quantum corrections to the general-relativistic setup and, eventually, some kind of instability.
Although the Universe is described by a real metric with Lorentzian
signature, energy-momentum tensors with an imaginary part and complex metrics
have been already considered, for example,
in Refs.~\cite{Halliwell:1989dy,Hartle:1983ai,Briscese:2022evf}.
Gibbons and Hawking~\cite{Gibbons:1976ue} showed that the Kerr metric
becomes complex-valued, and nondegenerate if the angular momentum
is assumed to be real, recovering the predicted thermodynamics underlying
the Kerr metric.
Later, Gibbons, Hawking, and Perry~\cite{Gibbons:1978ac} stated that the
path integral formalism of quantum gravity must be realized as the
infinite-dimensional analogue of a complex contour integration running
over complex spacetime metrics, based upon the fact that the action
of Euclidean quantum gravity has no positive-definite property.
They also showed that complex spacetimes can be employed in QFTs on
curved spacetimes and in quantum gravity, studying complex
extensions of the Kerr and Schwarzschild metrics.
Topological transitions were studied in Ref.~\cite{Louko:1995jw},
with complex spacetime metrics describing tunnelling trajectories.
In a semiclassical theory of gravity coupled to matter described by quantum fields, 
complex metrics can emerge.
Kontsevich and Segal classified the complex metrics in which a generic QFT
can be consistently coupled~\cite{Kontsevich:2021dmb}, which paved the way
for obtaining classes of suitable complex metrics for quantum gravity. 
The KS theorem establishes a criterion to determine which complex metrics
are compatible with the demand that QFTs may be consistently defined,
according to a bound on the summation of the arguments of its eigenvalues.
Witten showed that the KS criterion can be applied to
dynamical gravity~\cite{Witten:2021nzp}, by analyzing several complex
solutions and showing that the KS theorem selects a relevant set of complex metrics,
like complex black holes~\cite{Lehners:2021mah,Chen:2022hbi}. 
Visser~\cite{Visser:2021ucg} recently studied Feynman’s $i\,\varepsilon$-prescription
for propagators in QFT in terms of complex spacetime metrics, also extending this
prescription to QFT both on a fixed background and in a fuzzy spacetime geometry.
Besides proposing relevant extensions of the weak energy condition,
it implies constraints on the configuration space of admissible off-shell geometries
that are consistent with path integrals in quantum gravity~\cite{Visser:2021ucg,Andriolo:2021gen}. 
The 2-point correlation function of massive scalar fields 
can be also evaluated by semiclassical methods. Some spacelike  points cannot be connected by real geodesics, however complex
geodesics can link these points by analytical continuation to the sphere. Therefore 1-loop corrections to the correlator can be computed in  holographic models \cite{Chapman:2022mqd}.
%https://arxiv.org/pdf/2212.01398.pdf!!!!
Despite the fact that the quantum measure induces an imaginary contribution
to the energy-momentum tensor, we will show that such a contribution can be
compensated for in the interior region of the compact stellar distribution
and the metric remains real, given a particular mimic constraint. Other choices of the mimic constraint yields the 
metric with a complex radial component. 
Of course, this comes at the expense of the effective energy density and the radial
and tangential pressures.
\section{Gravitational decoupling and functional measure}
\label{s2}
We can now add the quantum correction described in the previous Section
to solutions obtained by applying the GD method~\cite{Ovalle:2017fgl}
by simply considering the field equations~\eqref{eom} with the matter energy-momentum 
tensor 
\begin{equation}
\label{emt}
T_{\mu\nu}
=
T^{\textsc{(m)}}_{\mu\nu}
+
\alpha\,\uptheta_{\mu\nu}
\ , 
\end{equation}
where
%{\color{red} WITH THE SIGNATURE  -+++  
%$T^\mu_{\ \nu}={\rm diag}(-\rho,p,p,p)$}
\begin{equation}
\label{perfect}
T^{\textsc{(m)}}_{\mu\nu}
=
(\rho +p)\,u_{\mu }\,u_{\nu }+p\,g_{\mu\nu}
\end{equation}
%https://www.maths.tcd.ie/~ipde/GR_Notes.pdf
represents the energy-momentum tensor of a perfect fluid
with 4-velocity $u^\mu$,  density $\rho$, and isotropic pressure $p$.
The term $\uptheta_{\mu\nu}$ in Eq.~(\ref{emt}) corresponds to the
contribution of additional sources, as recalled in the Introduction,
and it contains the parameter $\alpha$ so that the perfect fluid description
can be recovered in the limit $\alpha\to 0$.
Since the Einstein tensor in the left hand side of Eq.~\eqref{eom} satisfies the Bianchi identity,
the energy-momentum tensor~\eqref{emt} must satisfy the conservation equation
\begin{equation}
\nabla_\mu
T^{\mu\nu}
=
i\,\frac{\zeta}{2}\,
g^{\mu\nu}\,
\nabla_\mu
\left[
\log(1+\frac{\rho}{\Upsilon})
\right]
\simeq
i\,\bar\zeta\,g^{\mu\nu}\,
\nabla_\mu\,\rho
\ .
\label{dT0}
\end{equation}
The additional term of the effective energy-momentum tensor in the right-hand side of Eq. (\ref{eom1}) can be  interpreted as the components of $\theta_{\mu\nu}$ in Eq. (\ref{emt}). 
\par
A static and spherically symmetric metric can always be written 
in Schwarzschild-like coordinates as
\begin{equation}
\dd s^{2}
=
-
e^{\upnu (r)}\,\dd t^{2}
+
e^{\uplambda (r)}\,\dd r^{2}
+
r^{2}\,\dd\Omega^2
\ ,
\label{metric}
\end{equation}
where $\upnu =\upnu (r)$, $\uplambda =\uplambda (r)$, and
$\dd\Omega^2=\dd\theta^{2}+\sin^{2}\theta \,\dd\phi^{2}$.
The fluid inside the star has 4-velocity with
the only non-zero component $u^0=e^{-\upnu(r)/2}$
in the range $0\le r\le R_{\rm s}$, where $r=R_{\rm s}$
corresponds to the stellar surface.
The field equations~(\ref{eom}) for the metric~\eqref{metric} read
\begin{subequations}
\begin{eqnarray}
\label{ec1}
\kappa
\left[
\rho
+
i\,\frac{\zeta}{2}\, \log(1+\frac{\rho}{\Upsilon})
-
\alpha\,\uptheta_0^{\ 0}\right]
&\!=\!&
-\frac 1{r^2}
-
e^{-\uplambda }\left( \frac1{r^2}-\frac{\uplambda'}r\right)
\\
\label{ec2}
\kappa
\left[
p
-
i\,\frac{\zeta}{2}\,\log(1+\frac{\rho}{\Upsilon})
+
\alpha\,\uptheta_1^{\ 1}
\right]
&\!\!=\!\!&
\frac 1{r^2}
-
e^{-\uplambda }\left( \frac 1{r^2}+\frac{\upnu'}r\right)
\\
\label{ec3}
\kappa
\left[p
-
i\,\frac{\zeta}{2}\,\log(1+\frac{\rho}{\Upsilon})
+
\alpha\,\uptheta_2^{\ 2}
\right]
&\!\!=\!\!&
\frac {e^{-\uplambda }}{4}
\left(
\uplambda'\,\upnu'
+
2\frac{\upnu'-\uplambda'}{r}
-
2\,\upnu''
-
\upnu'^2
\right)
\ ,
\end{eqnarray}
\end{subequations}
where primes denote derivatives with respect to the areal radius $r$
and the conservation equation~\eqref{dT0} yields
\begin{equation}
p'
+
\frac{\upnu'}{2}\left(\rho+p\right)
+
\alpha\left(\uptheta_1^{\ 1}\right)'
-
\alpha\,\frac{\upnu'}{2}
\left(
\uptheta_0^{\ 0}-\uptheta_1^{\ 1}
\right)
+ 
\frac{2\,\alpha}{r}
\left(\uptheta_2^{\ 2}
-\uptheta_1^{\ 1}
\right)
=
i\,\frac{\zeta}{2}\,
\frac{\rho'}{\Upsilon+ \rho}
\ .
\label{con1}
\end{equation}
\par
From the system~\eqref{ec1}-\eqref{ec3}, one can define the effective density 
\begin{equation}
\mathring{\rho}
=
\rho
-
\alpha\,\uptheta_0^{\ 0}
+
i\,\frac{\zeta}{2}\,\log(1+ \frac{\rho}{\Upsilon}) 
\simeq
\rho
-
\alpha\,\uptheta_0^{\ 0}
+
i\,\bar\zeta\,\rho
\ ,
\label{efecden}
\end{equation}
whose imaginary part corresponds to the instability of the fluid
and measures the flow lifetime~\cite{Kuntz:2022kcw}. 
Also, the effective radial pressure can be read off as
\begin{equation}
\mathring{p}_{r}
=
p
+
\alpha\,\uptheta_1^{\ 1}
-
i\,\frac{\zeta}{2}\,\log(1+ \frac{\rho}{\Upsilon})
\simeq
p
+
\alpha\,\uptheta_1^{\ 1}
-
i\,\bar\zeta\,\rho
\ ,
\label{efecprera}
\end{equation}
as well as the effective tangential pressure
\begin{equation}
\mathring{p}_{t}
=
p
+
\alpha\,\uptheta_2^{\ 2}
-
i\,\frac{\zeta}{2}\,\log(1+ \frac{\rho}{\Upsilon})
\simeq
p
+
\alpha\,\uptheta_2^{\ 2}
-
i\,\bar\zeta\,\rho
\ .
\label{efecpretan}
\end{equation}
The additional source $\uptheta_{\mu\nu}$ induces the anisotropy  
\begin{equation}
\label{anisotropy}
\Pi(r,\alpha)
\equiv
\mathring{p}_{t}(r,\alpha)-\mathring{p}_{r}(r,\alpha)
=
\alpha\left(\uptheta_2^{\ 2}-\uptheta_1^{\ 1}\right)
\ ,
\end{equation}
which does not depend on the functional measure. 
\par
The MGD can be implemented to solve the system~\eqref{ec1}-\eqref{con1},
by considering the specific GR solution for an isotropic fluid described by
$T^{\textsc{(m)}}_{\mu \nu }$ in the limit $\alpha\to 0$,
which we write as
\begin{equation}
\dd s^{2}
=
-e^{\xi (r)}\,\dd t^{2}
+
\frac{\dd r^{2}}{\upmu(r)}
+
r^{2}\,\dd\Omega^{2}
\ ,
\label{pfmetric}
\end{equation}
where 
\begin{equation}
\label{standardGR}
\upmu(r)
\equiv
1
-
\frac{\kappa^2}{r}
\int_0^r
x^2\,\rho(x)\,\dd x
=
1
-
\frac{2\,m(r)}{r}
\ ,
\end{equation}
with $m$ the Misner-Sharp-Hernandez mass function representing
the energy within a sphere of areal radius $r$.
One can then switch on the parameter $\alpha$ to include the effects of the
source $\uptheta_{\mu\nu}$ on the perfect fluid solution.
The GD of the metric functions are then given by
\begin{subequations}
\begin{eqnarray}
\label{gd1}
\xi(r)
&\mapsto &
\upnu(r)
=
\xi(r)+\alpha\,\chi(r)
\ ,
\\
\label{gd2}
\upmu(r) 
&\mapsto &
e^{-\uplambda(r)}
=
\upmu(r)+\alpha\,f(r)
\ .
\end{eqnarray}
\end{subequations}
The MGD corresponds to setting $\chi=0$ and solving for $f\equiv f^\diamond$. 
The resulting metric is of the form in Eq.~\eqref{metric} with
\begin{eqnarray}
\label{expectg}
\upmu(r)\mapsto\,e^{-\uplambda(r)}
=
\upmu(r)+\alpha\,f^{\diamond}(r)
\ ,
\end{eqnarray}
whereas  $e^{\upnu(r)}$ is unaltered.
\subsection{MGD and functional measure}
Upon replacing Eq.~\eqref{expectg} in the field equations~\eqref{ec1}-\eqref{con1},
the system splits into two sets.
In the original (M)GD approach, the first set (corresponding to the limit $\alpha\to 0$)
is solved by the chosen metric~\eqref{pfmetric} by construction and one is left with a set
of equations that can be used to determine a consistent configuration of
$f^{\diamond}$ and $\uptheta_{\mu\nu}$. 
In the present case, beside $\uptheta_{\mu\nu}$, we have contributions from the
functional measure and we include their effects into the second set,
which we then solve perturbatively in $\bar\zeta\sim\zeta/\Upsilon$.  
\par
As we just mentioned, the first system of equations corresponds to the standard Einstein
field equations for the perfect fluid, that is
\begin{subequations}
\begin{eqnarray}
\label{ec1pf}
\kappa\,\rho
&\!=\!&
-\frac{1}{r^2}
-
\frac{\upmu}{r^2}
-
\frac{\upmu'}{r}
\\
\label{ec2pf}
\kappa\,p
&\!\!=\!\!&
\frac 1{r^2}
+\upmu\left( \frac 1{r^2}+\frac{\upnu'}r\right)
\\
\label{ec3pf}
\kappa\,p
&\!\!=\!\!&
-\frac{\upmu}{4}\left(2\,\upnu''+\upnu'^2+\frac{2\,\upnu'}{r}\right)
+\frac{\upmu'}{4}\left(\upnu'+\frac{2}{r}\right)
\ ,
\end{eqnarray}
\end{subequations}
along with Eq.~(\ref{con1}) in the limit $\alpha \to 0$,
\begin{equation}
\label{conpf}
p'
+
\frac{\upnu'}{2}\left(\rho+p\right)
=
0
\ .
\end{equation}
\par
The second set of equations contains the solution $\upnu$ of the
previous Eqs.~\eqref{ec1pf}-\eqref{conpf}, the MGD deformation $f^{\diamond}$
and the additional source $\uptheta_{\mu\nu}$, as well as the correction from the quantum
measure,
\begin{subequations}
\begin{eqnarray}
\label{ec1d}
\kappa
\left[
\uptheta_0^{\ 0}
+
i\,\bar\zeta\rho
\right]
&\!=\!&
-\frac{f^{\diamond}}{r^2}
-\frac{f^{\diamond'}}{r}
\\
\label{ec2d}
\kappa
\left[
\uptheta_1^{\ 1}
-
i\,\bar\zeta\rho\right]
&\!=\!&
f^{\diamond}\left(\frac{1}{r^2}+\frac{\upnu'}{r}\right)
\\
\label{ec3d}
\kappa
\left[
\uptheta_2^{\ 2}
-
i\,\bar\zeta\rho
\right]
&\!=\!&
-\frac{f^{\diamond}}{4}
\left(2\,\upnu''+\upnu'^2+\frac{2\,\upnu'}{r}\right)
+
\frac{f^{\diamond'}}{4}\left(\upnu'+\frac{2}{r}\right)
\ .
\end{eqnarray}
\end{subequations}
Furthermore, $\uptheta_{\mu\nu}$ must satisfy the conservation
equation~\eqref{con1} restricted to this sector, to wit 
\begin{equation}
\label{con1d}
\left(\uptheta_1^{\,\,1}\right)'
-\frac{\upnu'}{2}\left(\uptheta_0^{\ 0}-\uptheta_1^{\ 1}\right)
-\frac{2}{r}\left(\uptheta_2^{\ 2}-\uptheta_1^{\ 1}\right)
=
i\,\frac{\zeta}{2\,\alpha}\,\frac{\rho'}{\Upsilon+ \rho}
\ .
\end{equation}
The above implies that there is no direct exchange of energy-momentum
between the source $\uptheta_{\mu\nu}$ and the perfect fluid, 
but only with the quantum corrections which are on the same footing
as $\uptheta_{\mu\nu}$, so that the interaction between the two sectors
is purely gravitational.
\par 
As noticed in Ref.~\cite{Ovalle:2017fgl}, the right hand side of Eqs.~\eqref{ec1d}-\eqref{ec3d}
resemble the standard spherically symmetric field Eqs.~\eqref{ec1pf}-\eqref{ec3pf}
for $\zeta=0$, except for the missing $1/r^2$ terms.
This leads us to associate with $\uptheta_{\mu\nu}$ the effective energy density
$\mathring{\rho}$, effective radial pressure $\mathring{p}_r$, and effective tangential pressure
$\mathring{p}_t$, respectively given by
\begin{subequations}
\begin{eqnarray}
\label{theta1}
\mathring{\rho}
&\!=\!&
-\alpha\,{\uptheta^\diamond}_0^{\ 0}
=
-\alpha\,\uptheta_0^{\ 0}
-\frac{\alpha}{\kappa\,r^2}
-
i\,\bar\zeta\rho
\ ,
\\
\label{theta2}
\mathring{p}_r
&\!=\!&
\alpha\,{\uptheta^\diamond}_1^{\ 1}
=
\alpha\,\uptheta_1^{\ 1}
+
\frac{\alpha}{\kappa\,r^2}
+
i\,\bar\zeta\rho
\ ,
\\
\label{theta3}
\mathring{p}_t
&\!=\!&
\alpha\,{\uptheta^\diamond}_2^{\ 2}
=
\alpha\,\uptheta_2^{\ 2}
+
i\,\bar\zeta\rho
=
\alpha\,{\uptheta^\diamond}_3^{\ 3}
=
\alpha\,\uptheta_3^{\ 3}
+
i\,\bar\zeta\rho
\ .
\end{eqnarray}
\end{subequations}
Eq.~\eqref{con1d} then reads
\begin{equation}
\label{con1dd}
\left({\uptheta}_1^{\diamond\,1}\right)'
-\frac{\upnu'}{2}\left({\uptheta}_0^{\diamond\,0}-{\uptheta}_1^{\diamond\,1}\right)
-\frac{2}{r}\left({\uptheta}_2^{\diamond\,2}-{\uptheta}_1^{\diamond\,1}\right)
=
i\,\frac{\zeta}{\alpha}\,\frac{\rho'}{\Upsilon+ \rho}
.
\end{equation}
Since the conservation Eq.~(\ref{con1d}) [or \eqref{con1dd}] is again a linear combination of the
field Eqs.~\eqref{ec1d}-\eqref{ec3d}, the MGD eventually results in four unknown functions
$f^{\diamond}$, $\uptheta_0^{\ 0}$, $\uptheta_1^{\ 1}$, $\uptheta_2^{\ 2}$ satisfying 
Eqs.~\eqref{ec1d}-\eqref{ec3d} [or the equivalent anisotropic system~\eqref{theta1}-\eqref{theta3}].
\subsection{Compact objects}
A spherically symmetric star can be described by a perfect fluid of energy-momentum 
$T^{\textsc{(m)}}_{\mu \nu }$ localised within a radius $r=R_{\rm s}$, to which we can add  
both the (anisotropic) source $\uptheta_{\mu\nu}$ and the quantum correction from
the functional measure.
The interior geometry for $r<R_{\rm s}$ is therefore assumed to be described by the MGD metric 
\begin{equation}
\dd s^{2}
=
-e^{\upnu^{-}(r)}\,\dd t^{2}
+
\left[1-\frac{2\,\mathring{m}(r)}{r}\right]^{-1}
\dd r^2
+
r^{2}\dd\Omega^2
\ ,
\label{mgdmetric}
\end{equation}
where the interior mass function is given by
\begin{equation}
\label{effecmass}
\mathring{m}(r)
=
m(r)-\frac{r}{2}\,\alpha\,f^-(r)
\ , 
\end{equation}
with $m$ defined in Eq.~\eqref{standardGR} and $f^-=f^{\diamond}$ is the MGD introduced in
Eq.~\eqref{expectg} for $r<R_{\rm s}$.
\par
The metric~\eqref{mgdmetric} must match the outer geometry at $r=R_{\rm s}$,
which we write as
\begin{equation}
\dd s^{2}
=
-
e^{\upnu^{+}(r)}\,\dd t^{2}
+
e^{\uplambda^{+}(r)}\,\dd r^{2}
+
+r^{2}\dd\Omega^2
\ ,
\label{genericext}
\end{equation}
where $\upnu ^{+}(r)$ and $\uplambda ^{+}(r)$ are determined by the field equations~\eqref{eom}
for $r>R_{\rm s}$.
In particular, $T^{\textsc{(m)}}_{\mu \nu }=0$ and we also assumed the imaginary quantum correction
is negligible outside compact sources, which only leaves a possible contribution from $\uptheta_{\mu\nu}$
for $r>R_{\rm s}$.
The outer geometry will therefore be given by a MGD of the Schwarzschild metric,
\begin{equation}
\label{Schw}
\dd s^2
=
-
\left(1-\frac{2\,{\cal M}}{r}\right)\dd t^2
+
\left(1-\frac{2\,{\cal M}}{r}+\alpha\,f^+\right)^{-1}\dd r^2
+\dd\Omega^2
\ ,
\end{equation}
where $\cal M$ is the Arnowitt-Deser-Misner (ADM) mass of the system and the MGD $f^+=f^\diamond(r)$
is determined by the field Eqs.~\eqref{ec1d}-\eqref{ec3d} for $r>R_{\rm s}$.
\par
Continuity of the metric across the star surface implies
\begin{equation}
{\upnu ^{-}(R_{\rm s})}
=
{\upnu ^{+}(R_{\rm s})}
=
1-\frac{2\,\cal M}{R_{\rm s}}
\ ,
\label{ffgeneric1}
\end{equation}
and
\begin{equation}
1-\frac{2\,M_0}{R_{\rm s}}+\alpha\,f^-_{\rm s}
=
1-\frac{2\,{\cal M}}{R_{\rm s}}+\alpha\,f^+_{\rm s}
\ ,
\label{ffgeneric2}
\end{equation}
where $M_0=m(R_{\rm s})$ and we defined $F^\pm_{\rm s}\equiv \lim_{r\to\pm R_{\rm s}} F(r)$
for any function $F$.
\par
Continuity of the extrinsic curvature of the surface $r=R_{\rm s}$ along with the field
Eq.~\eqref{eom} imply 
\begin{equation}
p_{\rm s}
+\alpha\,(\uptheta_1^{\ 1})^{-}_{\rm s}
-
i\,\bar\zeta\rho
=
\alpha\,(\uptheta_1^{\ 1})^{+}_{\rm s}
\ ,
\label{matchingf}
\end{equation}
where $\rho_{\rm s}\equiv \rho^{-}(R_{\rm s})$ and
$p_{\rm s}\equiv p^{-}(R_{\rm s})$, since $\rho=p=0$ for $r>R_{\rm s}$.
Using Eq.~\eqref{ec2d} for the inner and outer geometries to eliminate
$\uptheta_1^{\ 1}$ then implies 
\begin{eqnarray}
\kappa\,p_{\rm s}
+
\frac{\alpha\,f_{\rm s}^-}{R_{\rm s}\left(R_{\rm s}-2\,\cal M\right)}
-i\,\kappa\,\bar\zeta\rho
=
\frac{\alpha\,f_{\rm s}^+}{R_{\rm s}\left(R_{\rm s}-2\,\cal M\right)}
\ .
\label{sfgenericf}
\end{eqnarray}
We note that the limit $\alpha=\zeta=0$ reproduces the standard condition
$p_{\rm s}=0$ for matching the isotropic fluid interior with an exact 
Schwarzschild metric in the exterior.
If the outer geometry is still given by the Schwarzschild metric with
additional sources and quantum corrections, $f^+=0$ and 
one might have a solid crust with $p_{\rm s}<0$~\cite{darkstars}. 
In this case, the imaginary quantum correction might induce an instability,
although we notice that this contribution vanishes at the surface if the density
$\rho_{\rm s}=0$.
\section{Minimal geometric deformation of Tolman~IV stars with quantum gravity corrections}
\label{mms5}
We can now apply the general formalism of Section~\ref{s2} to a particular solution of GR representing
a compact object.
Following Refs.~\cite{Ovalle:2013xla,Ovalle:2017wqi}, we shall consider the Tolman~IV
star of total mass $M_0$ and radius $R_{\rm s}$ satisfying the Buchdahl constraint for the compactness
$M_0/R_{\rm s}<4/9$, whose density and isotropic pressure are given by
\begin{subequations}
\begin{eqnarray}
%\label{tolmandensity}
\kappa\,\rho
&\!=\!&
\frac{3\,A^4\,M_0+A^2\left(3\,{R_{\rm s}^3}+7\,{M_0}\,r^2\right)+2\, r^2 \left({R_{\rm s}^3}+3 \,{M_0}\,r^2\right)}
{R_{\rm s}^3\left(A^2+2\,r^2\right)^2}
\\
%\label{tolmanpressure}
\kappa\,p
&=&
\frac{R_{\rm s}^3-M_0\left(A^2{{+}}3\,r^2\right)}{{R_{\rm s}^3}\left(A^2+2\,r^2\right)}
\ .
\end{eqnarray}
\end{subequations}
The corresponding interior metric, for $r<R_{\rm s}$, is of the form in Eq.~\eqref{pfmetric} with
\begin{subequations}
\begin{eqnarray}
\label{tolman00}
e^{\xi^-}
&\!=\!&
\left(1-\frac{3\,M_0}{R_{\rm s}}\right)\left(1+\frac{r^2}{A^2}\right)
=
e^{\upnu^-}
\\
\label{tolman11}
\upmu^-
&\!=\!&
\frac{A^2}{A^2+{2\,r^2}}\left(1-\frac{M_0\,r^2}{R_{\rm s}^3}\right)\left(1+\frac{r^2}{A^2}\right)
\ .
\end{eqnarray}
\end{subequations}
The constant $A$ can be expressed in terms of the unperturbed ADM mass $M_0$ by matching 
continuously the interior metric with the outer Schwarzschild geometry across $r=R_{\rm s}$,
which yields
\begin{equation}
A^2
=
\frac{R_{\rm s}^3}{M_0}\left(1-\frac{3\,M_0}{R_{\rm s}}\right)
\ .
\label{A}
\end{equation}
One then finds the metric functions
\begin{subequations}
\begin{eqnarray}
e^{\upnu^-}
&\!=\!&
1-\left(3-x^2\right) X
\label{tolman00x}
\\
\upmu^-
&\!=\!&
\frac{1-X\left[3-x^2\left(3-x^2\right) X\right]}{1-\left(3-2\,x^2\right) X}
\ ,
\label{tolman11x}
\end{eqnarray}
\end{subequations}
where we introduced the compactness $X\equiv M_0/R_{\rm s}$ and the
dimensionless radial coordinate $x=r/R_{\rm s}$ for convenience.
Likewise, we have 
\begin{subequations}
\begin{eqnarray}
\label{tolmandensityx}
\kappa\,R_{\rm s}^2\,\rho
&\!=\!&
3\,X\,\frac{2+X\left[2\,x^4\,X+x^2\left(3-7\,X\right)-9\left(1-X\right)\right]}
{\left[1-\left(3-2\,x^2\right) X\right]^2},
\\
\label{tolmanpressurex}
\kappa\,R_{\rm s}^2\,p
&=&
\frac{3\left(1{{-}}x^2\right) X^2}{1-\left(3-2\,x^2\right) X}
\ .
\end{eqnarray}
\end{subequations}
We remark that, according to the procedure described in Section~\ref{s2}, the metric 
functions $\upnu$ in Eq.~\eqref{tolman00x} and $\upmu$ in Eq.~\eqref{tolman11x}
represent the seed metric for the MGD satisfying Eqs.~\eqref{ec1pf}-\eqref{ec3pf} in which 
the density $\rho$ and isotropic pressure $p$ are given in Eqs.~\eqref{tolmandensityx}
and~\eqref{tolmanpressurex}.
\par
%\subsubsection{Mimic constraint I}
%\label{mmc1}
We next consider an MGD of the above solution which preserves the outer Schwarzschild
geometry~\eqref{Schw} with ${\cal M}=M_0$ and $f^+=0$. 
The matching Eq.~\eqref{matchingf}, can be solved by imposing the so-called mimic
constraint, now modified by the quantum correction, that is
\begin{equation}
\label{constraint3}
\uptheta_1^{\ 1}
=
p
+
i\,{\bar\zeta}\,\rho
\ .
\end{equation}
Following Eq.~\eqref{ec2pf}, the above can be expressed as~\footnote{We shall omit the superscripts
$\pm$ for simplicity when there is no confusion.}
\begin{equation}
\label{mmimic22}
\kappa\,\alpha\,R_{\rm s}^2\,\uptheta_1^{\ 1}
=
\frac{1}{x^2}
+\upmu
\left(\frac{1}{x^2}+\frac{\dot\upnu}{x}\right)\,, 
\end{equation}
where the dots represent derivatives with respect to $x$, that is $\dot F=R_{\rm s}\,F'$ for
any function $F$.
The interior MGD follows from Eq.~\eqref{ec2d} and reads
\begin{equation}
\alpha f^-
=
\upmu^-
+\frac{1}{1+x\,\dot\upnu^-}\left[1- {2i\,\bar\zeta}\,x^2\,R_{\rm s}^2\,\left(1-\frac{1}{\alpha}\right)\rho\right]
\ ,
\label{mimic3}
\end{equation}
implying that the interior metric encodes corrections due to quantum gravity and the
deformed metric element is given by
\beq
e^{-\uplambda^-}
&=&
2\upmu
+\,R_{\rm s}^2\left(\frac{1-3X+x^2\,X}{1-3X+3x^2\,X}\right)
\nonumber
\\
&&
+
{8\,i\,\bar\zeta}\,x^2\,R_{\rm s}^2\,\left\{1\!+\!\left(1\!-\!6\,X\right)^2\!\frac{X\,x^2}{1-3X}
\log\left[\left(1\!-\!3\,X\right)^2
\left(1\!+\!\frac{X\,x^2}{1-3X}\right)\right]\right\}\left(1-\frac{1}{\alpha}\right)
\rho\ ,
\label{tolman11dcc3}
\eeq
where again Eq.~\eqref{expectg} was employed additionally to Eq.~\eqref{tolman00x}. 
We now need to deal with the imaginary part of the metric function~\eqref{tolman11dcc3}. A complex effective action in the context of GR inherently evokes quantum aspects of this approach that might be interpreted classically in the context of GR. Although the imaginary term coming from the quantum measure cannot be arbitrariraly suppressed in the metric term \eqref{tolman11dcc3}, spacetime itself has a classical interpretation in GR and some kind of choice must be made hereon.
As we mentioned in the Introduction, a straightforward option which preserves the time
independence is given by taking the modulus of Eq.~\eqref{tolman11dcc3}. Since we are mainly interested in the effects induced by the quantum measure, 
we compute the norm of the metric term \eqref{tolman11dcc3} and subsequently Taylor-expand it in the running parameter $\bar\zeta\sim \zeta$ and just keep the leading
order, that is
\beq
\label{tolman11d3}
e^{-\uplambda^-}
&\!\simeq\!&
2\upmu
+R_{\rm s}^2\left(\frac{1-3X+x^2\,X}{1-3X+3x^2\,X}\right)\nonumber
\\
&&
+{4\frac{{\bar\zeta}^2}{\alpha^2}}\,x^4\,R_{\rm s}^4\,
\frac{\,\left\{1\!+\!\left(1\!-\!6\,X\right)^2\!\frac{X\,x^2}{1-3X}
\log\left[\left(1\!-\!3\,X\right)^2
\left(1\!+\!\frac{X\,x^2}{1-3X}\right)\right]\right\}^2}
{2\upmu
+
R_{\rm s}^2\left(\frac{1-3X+x^2\,X}{1-3X+3x^2\,X}\right)}
\left(\alpha-{1}\right)^2\rho^2
\ ,
\eeq
It is worth mentioning that the limit $\alpha\to1$ suppresses all quantum corrections
to the metric term, as the imaginary term in the metric component~\eqref{tolman11d3}
vanishes.
Also, if the MGD parameter $\alpha$ tends to zero, the quantum measure paramater
$\bar\zeta$ must go to zero faster, to avoid divergences in the imaginary term. 
\par 
It is instructive to write the metric component~\eqref{tolman11dcc3} in polar form
as $e^{-\uplambda^-}= \Theta\,e^{i\,\Phi}$, where
\beq
\Theta
&\!=\!\!&
\left\{
32 \frac{{\bar\zeta}^2}{\alpha^2} R_{\rm s}^4 x^4\left(\alpha-{1}\right)^2 \rho^2 \left(1+\frac{X
   (x-6 x X)^2}{1-3
   X} \log \left[(1-3 X)^2
   \left(\left(x^2-3\right) X+1\right)\right]\right)^2
\right.
\nonumber
\\
&&
\left.
\quad
+
\left[\frac{ R^2
   \left(\left(x^2-3\right) X+1\right)}{3
   \left(x^2-1\right) X+1}-\frac{ \left(X
   \left(\left(x^2-3\right) x^2
   X+3\right)-1\right)}{\left(2 x^2-3\right)
   X+1}\right]^2\right\}^{1/2}
\eeq
and~\footnote{The explicit form of $k=k(x,X)$ is not relevant for the following argument.}
\beq\label{atan3}
\Phi
&\!=\!&
\text{arctan}
\left[
-\frac{8 {\bar\zeta}  R_{\rm s}^2 x^2}{{\alpha}\,k(x,X)}\rho(x) \left(3
   \left(x^2-1\right) X+1\right) \left(\left(2
   x^2-3\right) X+1\right) \left(\alpha-{1}\right)^2\left(x^2 X (1-6 X)^2
   \right.
   \right.
   \nonumber
   \\
   &&
   \left.
   \left.
   \qquad
   \qquad
   \times\log
   \left((1-3 X)^2 \left(\left(x^2-3\right)
   X+1\right)\right)-3 X+1\right)\right]
   \ .
\eeq
Since only the metric component~\eqref{tolman11dcc3} is complex and $-\pi/2\le \arctan(b)\le \pi/2$,
for any argument $b\in\mathbb{R}$ (in particular the one in Eq. (\ref{atan3})), the MGD metric component
satisfies the criterion dictated by the KS theorem~\cite{Kontsevich:2021dmb}. 
\par
The MGD metric components can now be matched with the outer Schwarzschild solution~\eqref{Schw}
with $f^+=0$. 
Continuity expressed by Eqs.~\eqref{ffgeneric1} and \eqref{ffgeneric2} leads to the ADM mass
\beq
{\cal M}
&\!\simeq\!&
M_0
+
\frac{R_{\rm s}}{2}
\left(1\!-\!4X\right)
+
\frac{8\,{\bar\zeta}^2\left(\alpha-1\right)^2 R_{\rm s}^5}{\alpha^2\,(1-X)^2}
{\left\{1\!+\!\frac{X\left(1-6\,X\right)^2}{1-3X}
\log\left[\left(1\!-\!3\,X\right)
\left(1\!-\!2X\right)\right]\right\}^2}
\rho^2
\ ,
\qquad
\label{MASS3}
\eeq
where Eq.~\eqref{standardGR} was used.
The quantum gravity correction proportional to $\bar\zeta^2$ becomes comparable 
to the MGD for $|\bar\zeta|\simeq {|\alpha|}$.
Eq.~\eqref{constraint3} implies that the magnitude of the effective radial pressure
in Eq.~\eqref{efecprera} reads 
\beq
\mathring{p}_r
\!&\!=\!&\!
\frac{3X^2(1-x^2)}
{\kappa R_{\rm s}^2\left(1-3\,X+2\,x^2\,X\right)}
+
\frac{{\bar\zeta^2}\,v(x)\,X^2\left(\alpha\!-\!{1}\right)^2}
{\alpha^2\kappa\left(x^2 X\!-\!1\right)^4}\left[x^2 X \left(2 \left(x^2\!-\!3\right) X\!+\!3\right)\!+\!9X(X\!-\!1)
  \!+\!2\right]^2
  \ ,
\quad
\label{qpressrfc13}
\eeq
where 
\beq
v(x)
\equiv
\frac{(1-3\,X-2\,X\, x^2)(1-x^2)}{\kappa\left(1-3\,X+4\, X\, x^2\right)^2}
\left[1+\left(1-6\,X\right)^2\!\frac{X\,x^2}{1-3X}
\log\left[\left(1-3\,X\right)^2
\left(1+\frac{X\,x^2}{1-3\,X}\right)\right]\right]^2
\ .
\quad
\eeq
The radial pressure in Eq. \eqref{qpressrfc13} is real because the metric modulus is used.
It is worth mentioning that quantum gravity effects can be read off in the effective
radial pressure~\eqref{qpressrfc13} in the term accompanying the running parameter $\bar\zeta$.
\par
\begin{figure}[t]
\center
\includegraphics[scale=0.60]{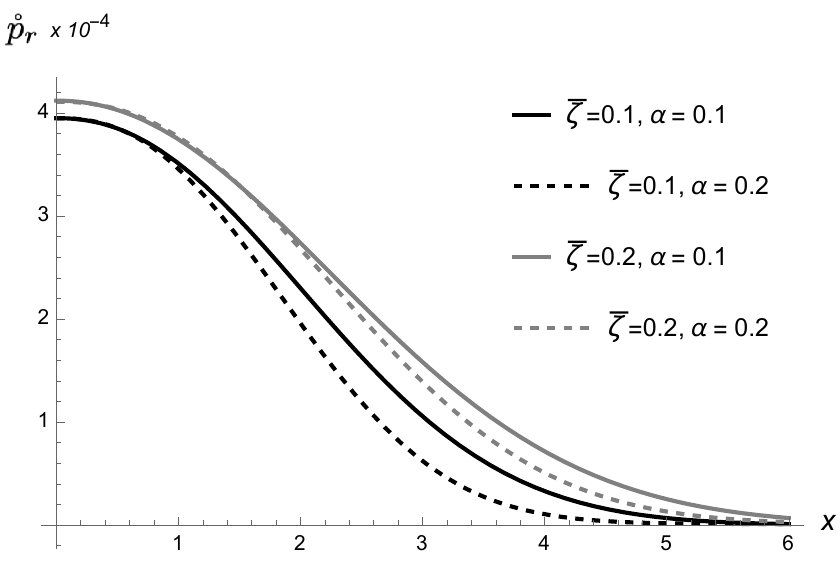}
\caption{\small Effective radial pressure $\mathring{p}_r$ of a stellar distribution of compactness $X=0.25$,
as a function of the radial coordinate, for different values of $\bar\zeta$ and $\alpha$.}
\label{gw7}      
\end{figure}

The effective radial pressure~\eqref{qpressrfc13} is displayed in Fig.~\ref{gw7} as a function of the radial coordinate,
for different values of the parameter $\bar\zeta$ governing quantum gravity corrections and also for two non-trivial
values of the GD charge $\alpha$. 
Quantum gravity effects implemented by the parameter $\bar\zeta$ in Eq.~\eqref{qpressrfc13}
show that the effective central radial pressure when $\bar\zeta =0.2$ is 4.3\% higher, when
compared to $\bar\zeta = 0.1$, for fixed values of the GD hairy charge.
For fixed values of $\bar\zeta$, the higher the GD hairy charge $\alpha$, the steepest the decrement
of the effective radial pressure is. It indicates that GD effects attenuate the 
effective radial pressure profile along the radial coordinate. 
\par
The effective density reads
\beq
\label{denfc}
\mathring{\rho}
&\!=\!&
\frac{X \left[2   X^3 \,x^2\left(3   X x^2+\!1\right)+(1\!-\!3   X)^2   X  \left(7   X \,x^2\!+\!3\right)+3 (1\!-\!3   X)^4
R_{\rm s}^4\right]}
{\kappa\left[2\,X^2 \,x^2+(1-3   X)^2 R_{\rm s}^2\right]^2} 
\nonumber
\\
&&
+\frac{6 \,\alpha    X  \left[  X \left(x^2\!-\!3\right)+1\right]}{R_{\rm s}^2\left[3   X (x^2\!-\!1)\!+\!1\right]^2}
+i \left(1-\frac{1}{\alpha}\right)\,{\bar\zeta}\,
w(x)\,
\frac{  X^2 v(x) (x^2\!-\!1)}{\left(4   X x^2\!-\!3   X\!+\!1\right)^2}
 \ ,
 \eeq 
where
\beq
w
=
\frac{9 \left(2 x^4-3 x^2+2\right)}{16 \kappa ^2 \left(x^2-1\right)^5 R_s^4} 
\left[x^6 (4 X-2)+x^4 (19-24 X)+x^2 (38 X-33)-22 X+20\right]
\ .
\eeq
As already mentioned, the imaginary part of the effective energy density in Eq.~\eqref{denfc}
corresponds to the instability of the degrees of freedom in the hydrodynamical fluid,
measuring the fluid lifetime.
This interpretation holds, in particular, for the case of the quark-gluon plasma, which is expected
to play a decisive role in the core of neutron and quark stars~\cite{Kuntz:2022kcw}. 
\par
\begin{figure}[t]
\center
\includegraphics[scale=0.60]{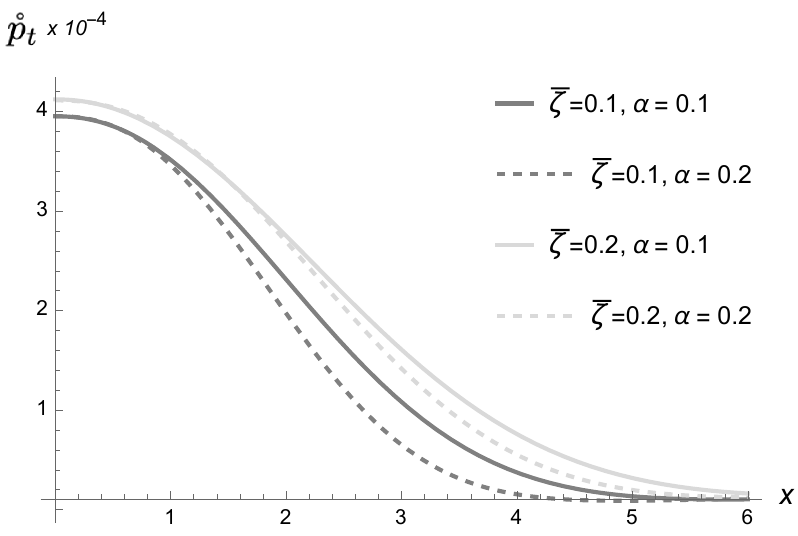}
\caption{\small Effective tangential pressure $\mathring{p}_t$ of a stellar distribution of compactness $X=0.25$,
as a function of the radial coordinate, for different values of $\bar\zeta$ and $\alpha$.}
\label{gw8}      
\end{figure}
\begin{figure}[H]
\center
\includegraphics[scale=0.60]{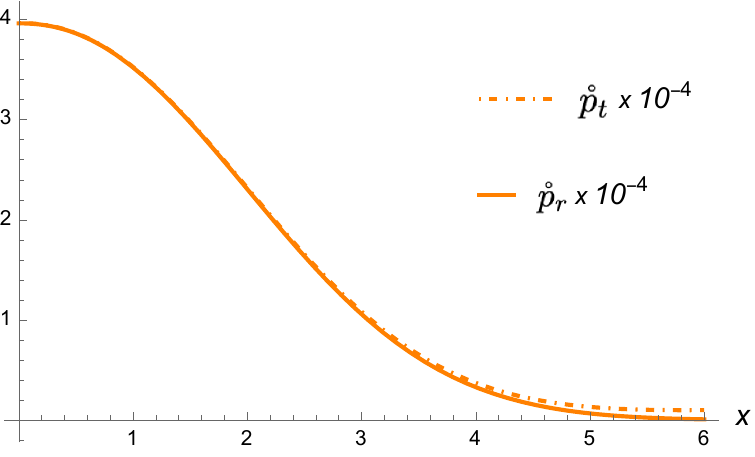}
\caption{\small Effective tangential and radial pressures of a stellar distribution of compactness $X=0.25$,
as a function of the radial coordinate, for $\bar\zeta =0.1$ and $\alpha = 0.1$.}
\label{gw7a}      
\end{figure}
\begin{figure}[H]
\center
\includegraphics[scale=0.60]{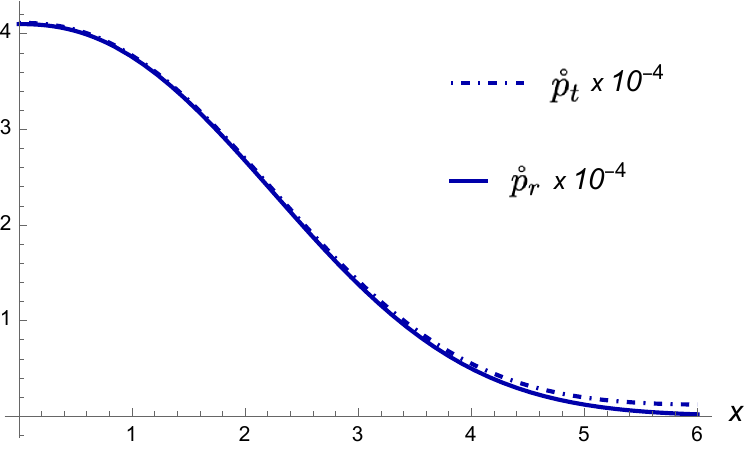}
\caption{\small Effective tangential and radial pressures of a stellar distribution of compactness $X=0.25$,
as a function of the radial coordinate, for $\bar\zeta =0.2 $ and $\alpha = 0.2$.}
\label{gw8a}      
\end{figure}

The effective tangential pressure, after Taylor-expanding it in terms of the quantum gravity parameter
$\zeta$ up to fourth order, reads
\beq
\label{pretanft3}
\mathring{p}_t
&\!\simeq\!&
\frac{3X^4(1\!-\!x^2)}{\kappa\,R_{\rm s}^2\left(1-3X+2x^2X\right)}
+\frac{X\,x^2}{{\kappa   (x^2-1)+1}}
\nonumber
\\
&&
+
\frac{2\bar\zeta^2}{\alpha^2}  w(x)\frac{X\kappa\,v^2(x)R_{\rm s}^2 (x^2-1)
\left(\alpha-{1}\right)^2}{3\left(4 X x^2-3 X +1\right)^2}{({2 X x^2 -1+3 X)}}
\ .
\eeq 
Fig.~\ref{gw8} shows this effective tangential pressure as a function of the radial coordinate,
for several values of the parameter $\bar\zeta$.
The anisotropic factor is illustrated in Figs.~\ref{gw7a} and~\ref{gw8a} as a function of the radial coordinate,
for different values of $\bar\zeta$ and $\alpha$.
Those graphs show that the anisotropy increases towards the stellar surface.
\par
\begin{figure}[t]
\center
\includegraphics[scale=0.70]{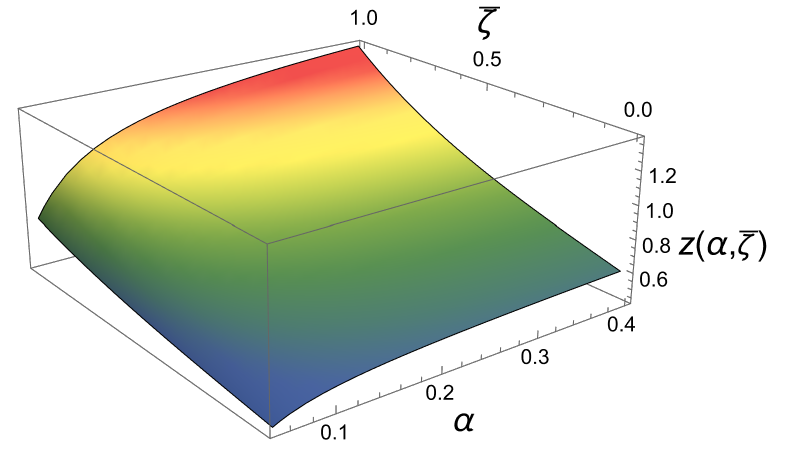}
\caption{\small Anisotropic surface redshift of a stellar distribution of compactness $X=0.25$,
as a function of the GD hairy parameter $\alpha$ and the quantum gravity parameter $\bar\zeta$. }
\label{gw123}      
\end{figure}

An interesting quantity to consider is the surface redshift, which depends on both $\bar\zeta$ and $\alpha$
and can be expressed as
\beq
\label{redss}
z(\alpha, \zeta)
=
\frac{1}{\sqrt{1-\frac{2\mathcal{M}(\alpha,\bar\zeta)}{R_{\rm s}}}}-1
\ .
\eeq
It can be directly obtained from the ADM mass in Eq.~\eqref{MASS3} and is displayed in Fig.~\ref{gw123},
as a function of the MGD charge $\alpha$ and the quantum gravity parameter $\zeta$.
The anisotropic factor amplifies the gravitational redshift at the stellar surface.
Therefore, for each fixed value of $\zeta$, a distant observer detects a more compact stellar
distribution for $\alpha>0$, when compared to the isotropic case.
Reciprocally, for each fixed value of the GD charge $\alpha$, a distant observer sees a stellar distribution
that is more compact, for $\bar\zeta>0$.
When $\bar\zeta \to0$ the redshift is a function of the hairy charge $\alpha$ only and reproduces
the result in Ref.~\cite{Ovalle:2017wqi}. 
The larger the magnitude of quantum gravity effects driven by $\bar\zeta$, the bigger the surface redshift.
These features comply with Eq.~\eqref{MASS3}, which in particular states that $\mathcal{M}>M_0$.
They are also compatible with the recent bounds for the surface redshift in realistic anisotropic stellar models. 
\par
\begin{figure}[t]
\center
\includegraphics[scale=0.60]{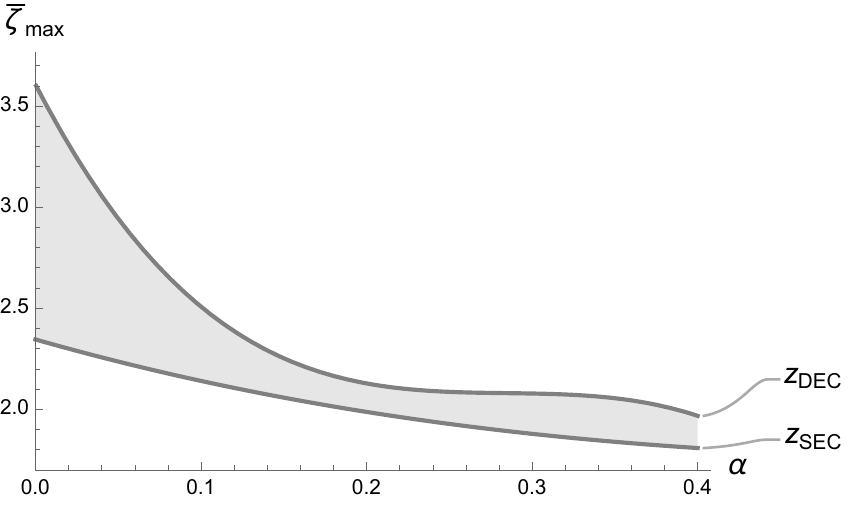}
\caption{\small Upper bounds $\bar\zeta_{\rm max}$  
as a function of $\alpha$, for $10^{-5}\leq \alpha\leq 0.4$, according to the bounds $z_{\rm SEC} = 3.840$
and  $z_{\rm DEC} = 5.211$ on the surface redshift~\cite{Ivanov:2002xf}. }
\label{gw10}      
\end{figure}

The upper limit $z=5.211$ was obtained for compact stellar distributions that satisfy dominant energy
conditions (DEC) in Ref.~\cite{Ivanov:2002xf}.
It yields the upper bounds $\bar\zeta_{\rm max} = 2.18\, $, when $\alpha = 0.4$,  and
$\bar\zeta_{\rm max} = 2.268$, for $\alpha=0.2$.
The upper limit $z=3.840$, for strong energy conditions (SEC) yield, apiece, the upper bounds
$\bar\zeta_{\rm max} = 2.088$, when $\alpha = 0.4$,  and $\zeta_{\rm max} = 2.009$, for $\alpha=0.2$.
More generally, the upper bounds $\zeta_{\rm max}$, with $\alpha$ varying, are plotted in Fig.~\ref{gw10}.
Higher values of $\alpha$ lower the maximum value $\bar\zeta_{\rm max}$.
We remark the fact that $\alpha$ and $\bar\zeta$ are completely independent, as they respectively represent
the MGD and the functional measure as different sources.
Fig.~\ref{gw10} illustrates the fact that the upper bound on the redshift implies an upper bound $\bar\zeta_{\rm max}$
on the parameter $\bar\zeta$, for each fixed value in the range  $10^{-5}\leq \alpha\leq 0.4$ considered.
Higher magnitudes of the GD hairy charge $\alpha$ induce suppression of $\bar\zeta_{\rm max}$.
The numerical results can be interpolated by the polynomial curve,  
\beq
\bar\zeta^{\scalebox{.55}{\textsc{DEC}}}_{\rm max}
&\!\simeq\!&
-65.077 \alpha^3+ 55.544  \alpha^2 - 15.907 \alpha +3.7076
\ ,
\eeq
for the upper bound $z_{\rm DEC} = 5.211$ on the surface redshift, and by 
\beq
\bar\zeta^{{\scalebox{.55}{\textsc{SEC}}}}_{\rm max}
&\!\simeq\!&
-1.1329\alpha^3 +2.9280\alpha^2 -2.3338 \alpha + 2.3448
\ ,
\eeq
for the upper bound $z_{\rm SEC} = 3.840$, both with interpolation errors $<10^{-4}$.
\par
Different choices for the mimic constraints are analysed in Appendix~\ref{appA}.
\section{conclusions}
\label{secc}
We have studied the influence of the functional measure in the 1-loop effective action
of quantum gravity within the GD approach.
The most striking feature induced by the functional measure is the addition of imaginary terms
in the effective energy-momentum tensor that sources the Einstein field equations.
One generally expects that such terms reflect inherent instabilities.
\par
For a suitable choice of the mimic constraint to preserve the outer Schwarzschild vacuum,
we showed that it naturally generates complex metric functions that satisfy the KS criterium,
showing the compatibility between the metric solutions obtained and the QFT side of quantum gravity. 
Effective field equations for compact stellar objects were therefore implemented
in the context of both the GD and quantum gravity to generate consistent solutions
for the interior of a compact object.
The role of both the GD and quantum gravity on the profiles of the effective radial and
tangential pressures, as well as on the effective energy density, was analysed
for compact self-gravitating anisotropic stellar distributions obtained from the Tolman~IV
family of solutions.
\par
For a different choice of the mimic constraint detailed in Appendix~\ref{app1},
the GD method coupled with the additional imaginary term due to the functional measure
can still accommodate real metrics, only affecting the hidden sector
represented by the GD term in the energy-momentum in the Einstein equations.
It is a consequence of the modified mimic constraint and the cancellation of the
functional measure contribution in the outer region of the stellar distribution. 
\par
Finally, we studied the absolute bounds on the surface redshift of a stellar distribution. 
Other mimic constraints and extensions considered in Ref.~\cite{Ovalle:2017wqi}
can be also employed in the context of the functional measure.
The upper bounds $\bar\zeta_{\rm max}$ are then higher, showing that the mimic constraint
for the pressure provides the most stringent upper limits to the parameter regulating quantum
gravity effects. 
\subsubsection*{Acknowledgments} 
R.C.~is partially supported by the INFN grant FLAG and his work has also been carried
out in the framework of activities of the National Group of Mathematical Physics (GNFM, INdAM). 
I.K.~thanks the National Council for Scientific and Technological Development -- 
CNPq (Grant No.~303283/2022-0 and Grant No.~401567/2023-0) for financial support.
R.d.R.~thanks to The S\~ao Paulo Research Foundation -- FAPESP
(Grants No. 2021/01089-1 and No.~2024/05676-7),  
CNPq (Grants No. 303742/2023-2 and No. 401567/2023-0), and the Coordination for the Improvement of
Higher Education Personnel (CAPES-PrInt~88887.897177/2023-00), for partial financial
support.
R.dR.~thanks R.C.~and DIFA, Universit\`a di Bologna, for the hospitality. 
\appendix
\section{Reality conditions and the absolute value}
\label{app:abs}
{\color{black}
There could be some freedom in choosing the reality condition $f: U\subset \mathbb{C}\to \mathbb{R}$. However, the absolute value can be obtained under very minimal and reasonable assumptions: (i) basis-independence of $\mathbb{C}$ and (ii) both the complex quantity and its real counterpart should scale in the same way.

Regarding (i), there is no dispute that the result of $f(z)$ should not depend on the choice of coordinates for the complex plane, namely whether one writes e.g. $z=\Re(z)+i \, \Im(z)$ or $z = r e^{i \theta}$. The only scalars in $\mathbb{C}$ are functions of $z \bar z$, thus $f(z) = g(z \bar z)$.

The second assumption is just a statement about the scaling of units. For example, if we change the units of the complex energy density from eV$^4$ to GeV$^4$, the real (and physical) energy density would scale along. It is a fair physical assumption. This implies that $f$ is a homogenous function of first degree, namely $f(\lambda z) = \lambda f(z)$, or 
\[
g(\lambda^2 \, z \bar z) = \lambda g(z \bar z)
\ .
\]
Differentiating this equation with respect to $\lambda$ gives:
\[
2\, \lambda \, z \bar z \, g'(\lambda^2 \, z \bar z) =  g(z \bar z)
\ .
\]
Since this holds for any $\lambda$, we can take $\lambda=1$ and solve the differential equation, which results in:
\[
g(z \bar z) \sim |z|
\ .
\]
This justifies the use of the absolute value when imposing reality conditions.
We therefore define ``physical quantities'' by the absolute value of the complex ones.
}

\section{Alternative mimic constraints}
\label{appA}
We provide here details about different implementations of the mimic constraint that 
preserve the outer Schwarzschild metric.
\subsection{Case~I}
\label{app1}
We note that the matching condition in Eq.~\eqref{matchingf} can also be solved by imposing
\begin{equation}
\label{constraint}
\alpha\,\uptheta_1^{\ 1}
=
p
+
i\,\bar\zeta\rho
\ ,
\end{equation}
which, according to Eq.~\eqref{ec2pf}, can be written as
\begin{equation}
\label{mmimic2}
\kappa\,\alpha\,R_{\rm s}^2\,\uptheta_1^{\ 1}
=
\frac{1}{x^2}
+\upmu
\left(\frac{1}{x^2}+\frac{\dot\upnu}{x}\right)
+
i\,\kappa\,R_{\rm s}^2\,\bar\zeta\rho\ ,
\end{equation}
The interior MGD deformation is then determined by Eq.~\eqref{ec2d}. 
The imaginary contribution due to the functional measure drops out and the metric remains real, as 
\begin{equation}
\alpha\,f^-
=
\upmu^-
+\frac{1}{1+x\,\dot\upnu^-}
\ ,
\label{mimic}
\end{equation}
which yields
\beq
e^{-\uplambda^-}
=
\upmu^-+\alpha\,f^-
=
2\,\upmu^-
+
\frac{1}{1+x\,\dot\upnu^-}
\ .
\eeq
Continuity expressed by Eqs.~\eqref{ffgeneric1} and \eqref{ffgeneric2} leads to the ADM mass
\begin{eqnarray}
\label{fff1}
{\cal M}
=
\frac{R_{\rm s}}{2}=M_0.
\eeq
This case is therefore rather trivial and no modifications from both the GD hairy charge and the parameter
encoding quantum gravity effects are obtained. 
\subsection{Case~II}
\label{app2}
Let us finally consider
\begin{equation}
\label{constraint2}
\alpha\,\uptheta_1^{\ 1}
=
p
\ ,
\end{equation}
which, following Eq.~\eqref{ec2pf}, can be expressed as
\begin{equation}
\label{mmimic22}
\kappa\,\alpha\,R_{\rm s}^2\,\uptheta_1^{\ 1}
=
\frac{1}{x^2}
+\upmu
\left(\frac{1}{x^2}+\frac{\dot\upnu}{x}\right)\ .
\end{equation}
The interior MGD comes from Eq.~\eqref{ec2d}, that is
\begin{equation}
\alpha f^-
=
\upmu^-
+\frac{1}{1+x\,\dot\upnu^-}\left(1- {2i\,\bar\zeta}\,x^2\,R_{\rm s}^2\,\rho\right)
\ ,
\label{mimic2}
\end{equation}
so that
\beq
e^{-\uplambda^-}
&\!=\!&
2\,\upmu
+\,R_{\rm s}^2\left(\frac{1-3X+x^2\,X}{1-3X+3x^2\,X}\right)
\nonumber
\\
&&
+
{8\,i\,\bar\zeta}\,x^2\,R_{\rm s}^2\,\left\{1\!+\!\left(1\!-\!6\,X\right)^2\!\frac{X\,x^2}{1-3X}
\log\left[\left(1\!-\!3\,X\right)^2
\left(1\!+\!\frac{X\,x^2}{1-3X}\right)\right]\right\}
\rho\ ,
\label{tolman11dcc2}
\eeq
where again Eq.~\eqref{expectg} and~\eqref{tolman00x} were employed. 
\par
We again deal with the imaginary part of the metric function~\eqref{tolman11dcc2} by taking its
modulus.
By Taylor-expanding in the running parameter $\zeta$ and keeping the leading order, we obtain
\beq
\label{tolman11d2}
e^{-\uplambda^-}
&\!\simeq\!&
2\upmu
+R_{\rm s}^2\left(\frac{1-3X+x^2\,X}{1-3X+3x^2\,X}\right)\nonumber
\\
&&
+{4{\bar\zeta}^2}\,x^4\,R_{\rm s}^4\,
\frac{\,\left\{1\!+\!\left(1\!-\!6\,X\right)^2\!\frac{X\,x^2}{1-3X}
\log\left[\left(1\!-\!3\,X\right)^2
\left(1\!+\!\frac{X\,x^2}{1-3X}\right)\right]\right\}^2}
{2\upmu
+
R_{\rm s}^2\left(\frac{1-3X+x^2\,X}{1-3X+3x^2\,X}\right)}
\rho^2
\ ,
\eeq
The polar form $e^{-\uplambda^-}= \Theta\,e^{i\,\Phi}$ is now given by
\beq
\Theta
&\!=\!\!&
\left\{
32 {\bar\zeta}^2 R_{\rm s}^4 x^4 \rho^2 \left(1+\frac{X
   (x-6 x X)^2}{1-3
   X} \log \left[(1-3 X)^2
   \left(\left(x^2-3\right) X+1\right)\right]\right)^2
\right.
\nonumber
\\
&&
\left.
\quad
+
\left[\frac{ R^2
   \left(\left(x^2-3\right) X+1\right)}{3
   \left(x^2-1\right) X+1}-\frac{ \left(X
   \left(\left(x^2-3\right) x^2
   X+3\right)-1\right)}{\left(2 x^2-3\right)
   X+1}\right]^2\right\}^{1/2}
\eeq
and
\beq\label{atan}
\Phi
&\!=\!&
\text{arctan}
\left[
-\frac{8 \bar\zeta  R_{\rm s}^2 x^2}{k(x,X)}\rho(x) \left(3
   \left(x^2-1\right) X+1\right) \left(\left(2
   x^2-3\right) X+1\right) \left(x^2 X (1-6 X)^2\right.\right.\nonumber\\ &&\left.\left.\qquad\qquad\times\log
   \left((1-3 X)^2 \left(\left(x^2-3\right)
   X+1\right)\right)-3 X+1\right)\right],
\,
\eeq
where 
\beq
k(x,X) &=& (3 X-1)
   \left[\left(x^2-3\right) X+1\right] \left[\left(2 x^2-3\right) X-3 x^2 \left(x^2-1\right) X^2+1\right].
\eeq
The MGD metric still satisfies the criterion established in the KS theorem~\cite{Kontsevich:2021dmb}. 
\par
The ADM mass $\cal M$ can be expressed as
\beq
{\cal M}
\simeq
M_0
+
\frac{R_{\rm s}}{2}\,
\left(1\!-\!4X\right)
+
\frac{8\,{\bar\zeta}^2\,R_{\rm s}^5}{(1-X)^2}
{\left\{1\!+\!\frac{X\left(1-6\,X\right)^2}{1-3X}
\log\left[\left(1\!-\!3\,X\right)
\left(1\!-\!2X\right)\right]\right\}^2}
\rho^2\ ,
\label{MASS2}
\eeq
where Eq.~(\ref{standardGR}) was used.
\par
Eq.~\eqref{constraint} implies that the magnitude of the effective radial pressure
$\mathring{p}_r=\mathring{p}_r(x,\zeta)$ in Eq.~\eqref{efecprera} reads 
\beq
\mathring{p}_r
=
\frac{3\,X^2\,(1-x^2)}
{\kappa R_{\rm s}^2\left(1-3\,X+2\,x^2\,X\right)}
+
{{\bar\zeta^2}\,v(x)\,X^2}\,
\frac{\left[x^2 X \left(2 \left(x^2-3\right) X+3\right)+9X(X-1)
  +2\right]^2}{\kappa\left(x^2 X-1\right)^4}
  \ .
\label{qpressrfc1}
\eeq
The effective radial pressure~\eqref{qpressrfc1} is depicted in Fig.~\ref{gw1}
as a function of the radial coordinate, for different values of the parameter $\zeta$
governing quantum gravity corrections. 
Quantum gravity effects implemented by the parameter $\bar\zeta$ in Eq.~\eqref{qpressrfc1}
show that the effective central radial pressure when $\bar\zeta =0.1$ is 11.3\% smaller than when
$\bar\zeta = 0$, whereas for $\zeta = 0.2$, this deviation increases up to $19.9\%$.
Although quantum gravity effects make the central radial pressure decrease,
for $x\gtrsim6$ these values equalize.
Fig.~\ref{gw1} also shows that in the range $1.8\lesssim x\lesssim 2.4$ the values
of the effective radial pressure are essentially the same.
\begin{figure}[t]
\center
\includegraphics[scale=0.60]{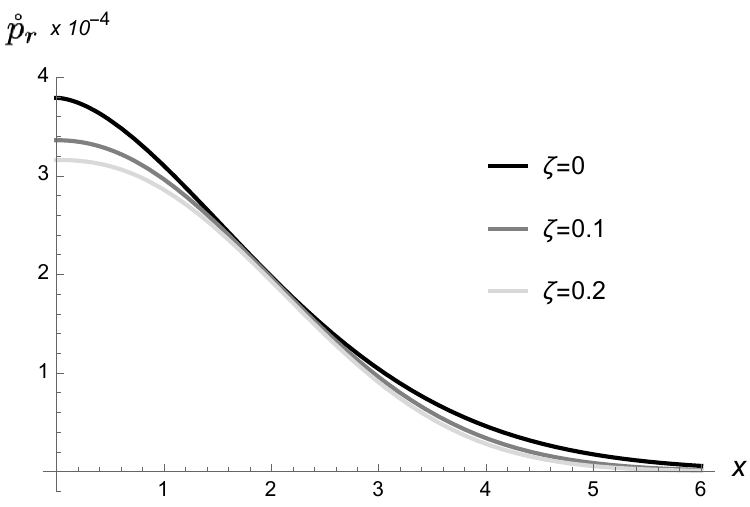}
\caption{\small Effective radial pressure $\mathring{p}_r$ of a stellar distribution of compactness $X=0.25$,
as a function of the radial coordinate, for different values of $\bar\zeta$.}
\label{gw1}      
\end{figure}
\par
The effective tangential pressure, after Taylor-expanding it in terms of the quantum gravity
parameter $\bar\zeta$ up to $\mathcal{O}(\bar\zeta^4)$, reads
\beq
\label{pretanft}
\mathring{p}_t(x,\bar\zeta)
&\!=\!&
\frac{3X^4(1\!-\!x^2)}{\kappa\,R_{\rm s}^2\left[1-3X+2x^2X\right]}
+
\frac{X\,x^2}{{\kappa   (x^2-1)+1}}
\nonumber
\\
&&
+ {2\,\bar\zeta^2}  w(x)\,
\frac{X\kappa\,v^2(x)R_{\rm s}^2 (x^2-1)\left[{({2 X x^2 -1+3 X)}}\right]}{3\left(4 X x^2-3 X +1\right)^2}
\ .
\eeq 
Fig.~\ref{gw3} shows the effective tangential pressure $\mathring{p}_t$ profile \eqref{pretanft}
as a function of the radial coordinate, for several values of the parameter $\zeta$.
Quantum gravity effects induce the central tangential pressure also to decrease,
although at a lower rate, compared to the effective radial pressure.
For $r\gtrsim6.0$ these values also equalize.
Fig.~\ref{gw3} shows that in the range $1.9\lesssim x\lesssim 2.3$ the values of the effective
tangential pressure attain very similar values.
%
%\blt{All the equations for the metric component, the ADM mass and the effective radial
%and tangential pressures can be obtained from the ones in Subsec. \ref{mmc1} in the formal limit $\alpha\to\infty$. }
%
\begin{figure}[b]
\center
\includegraphics[scale=0.60]{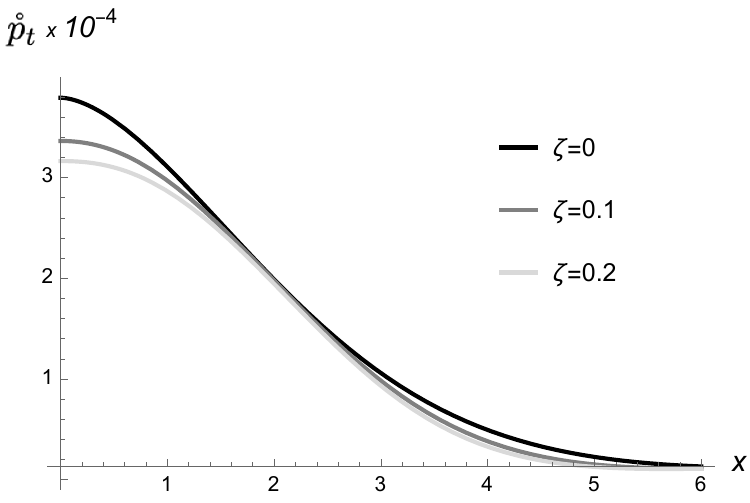}
\caption{\small Effective tangential pressure $\mathring{p}_t$ of a stellar distribution of compactness
$X=0.25$, as a function of the radial coordinate, for different values of $\zeta$.}
\label{gw3}      
\end{figure}
\par
The anisotropic factor is illustrated in Figs.~\ref{gw4}-\ref{gw6} as a function of the radial coordinate,
for different values of $\bar\zeta$, as the difference between the effective tangential and radial pressures
of a stellar distribution of compactness $X=0.25$, for different values of $\bar\zeta$.
One can see that the anisotropy factor increases towards the stellar surface.
The effective tangential pressure decreases as a function of the radial coordinate, however at a lower rate,
compared to the effective radial pressure.
For $x\gtrsim2.0$ the difference between the effective radial and the tangential  pressures
is already noticeable and attains a maximum value at $x\sim 6.0$. 
\begin{figure}[t]
\center
\includegraphics[scale=0.60]{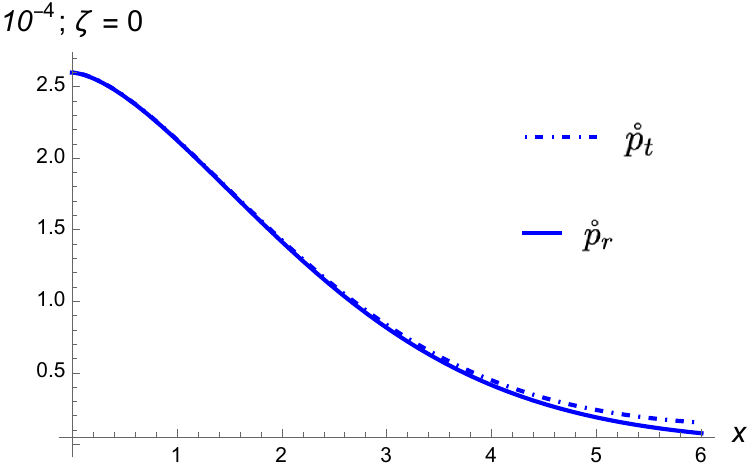}
\caption{\small Effective tangential and radial pressures of a stellar distribution of compactness $X=0.25$,
as a function of the radial coordinate, for $\bar\zeta =0$.}
\label{gw4}      
\end{figure}
\begin{figure}[t]
\center
\includegraphics[scale=0.60]{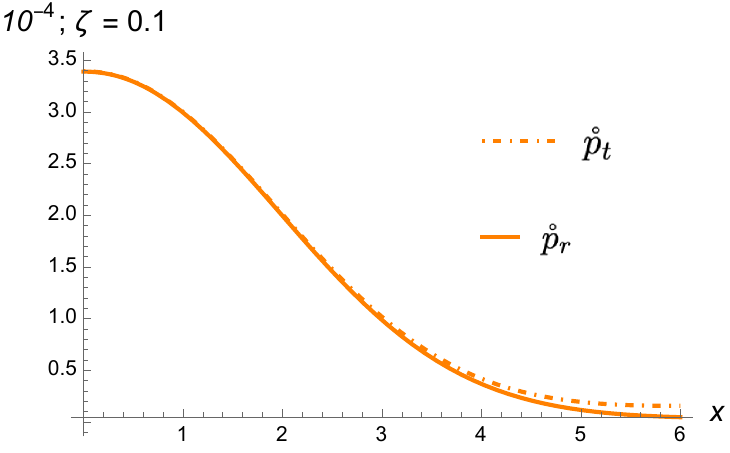}
\caption{\small Effective tangential and radial pressures of a stellar distribution of compactness
$X=0.25$, as a function of the radial coordinate, for $\bar\zeta =0.1 $.}
\label{gw5}      
\end{figure}
\begin{figure}[t]
\center
\includegraphics[scale=0.60]{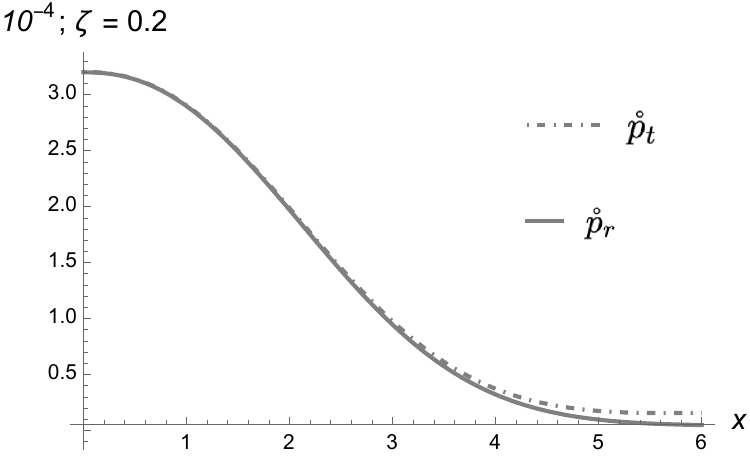}
\caption{\small Effective tangential and radial pressures of a stellar distribution of compactness $X=0.25$,
as a function of the radial coordinate, for $\bar\zeta =0.2 $.}
\label{gw6}      
\end{figure}
\par
The surface redshift~\eqref{redss} is displayed in Fig.~\ref{gw2}, as a function of the quantum gravity
parameter $\bar\zeta$.
Anisotropy is shown to amplify the gravitational redshift at the stellar surface.
Therefore, for each fixed value of $\bar\zeta$, a distant observer detects a more compact stellar
distribution, when compared to the isotropic case corresponding to $\bar\zeta\to0$,
where no quantum gravity effects set in.
The larger the magnitude of quantum gravity effects driven by $\bar\zeta$, the bigger the surface redshift is.
These features comply with Eq.~\eqref{MASS2}, which in particular states that $\mathcal{M}>M_0$.
They are also compatible with the recent bounds for the surface redshift in realistic anisotropic stellar models. 
\par
The upper limit $z=5.211$ was obtained for the redshift of compact stellar distributions that satisfy dominant
energy conditions (DEC) in Ref.~\cite{Ivanov:2002xf} and yield, the upper bound $\zeta_{\rm max} = 2.373$.
The redshift upper limit $z=3.840$, for strong energy conditions (SEC) yields the upper bound
$\bar\zeta_{\rm max} = 2.072$. 
\begin{figure}[t]
\center
\includegraphics[scale=0.60]{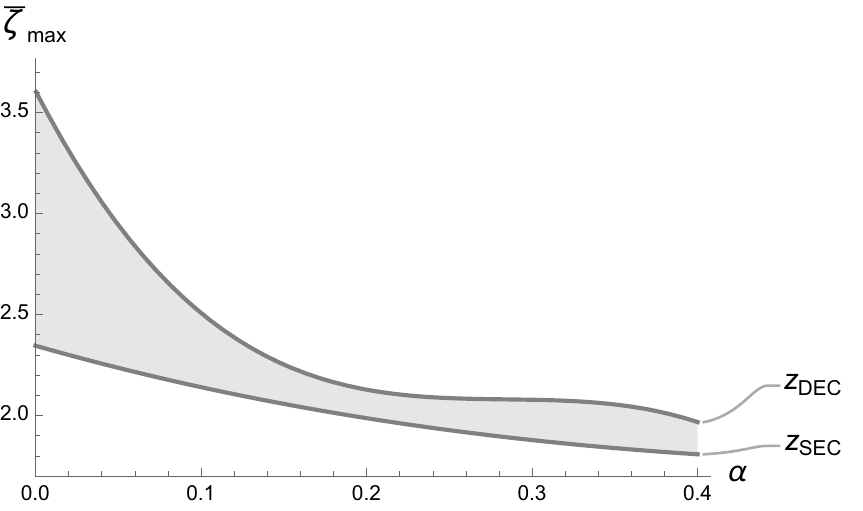}
\caption{\small Anisotropic surface redshift of a stellar distribution of compactness $X=0.25$,
as a function of the quantum gravity parameter $\bar\zeta$. }
\label{gw2}      
\end{figure}

\paragraph*{Data Availability Statement:} No Data associated in the manuscript.
\bibliography{bib_GD_QG.bib}
\end{document}